\documentclass[final,1p,times]{elsarticle}

\usepackage{amssymb}
\usepackage{amsmath}
\usepackage{epstopdf}
\usepackage{float}
\usepackage{bm}
\usepackage{color,soul}
\usepackage{hhline,colortbl}
\usepackage[colorlinks=true,linkcolor=black, citecolor=blue, 
urlcolor=blue]{hyperref}
\usepackage{flowchart}
\usepackage{tikz}
\usepackage{textcomp}
\usetikzlibrary{shapes.geometric, arrows,calc, shapes.misc, matrix, shapes, arrows, calc, positioning, shapes.geometric, shapes.symbols, shapes.misc, intersections, chains, scopes}
\tikzstyle{startstop} = [rectangle, rounded corners, minimum width=3cm, minimum height=1cm,text
centered, draw=black, fill=red!30]
\tikzstyle{io} = [trapezium, trapezium left angle=70, trapezium right angle=110, minimum 
width=3cm, minimum height=1cm, text centered, text width=3cm,draw=black, fill=blue!30]
\tikzstyle{process} = [rectangle, minimum width=3cm, minimum height=1cm, text centered, 
draw=black, fill=orange!30]
\tikzstyle{decision} = [diamond,  aspect=3.5, minimum width=3cm, minimum height=1cm, text centered, text width=4cm, draw=black, fill=green!30]
\tikzset{
    loop/.style={ 
        draw,
        chamfered rectangle,
        chamfered rectangle xsep=2cm,text centered,
        text width=10em,draw=black, fill=yellow!30
    },
}
\tikzstyle{subroutine} =[draw, predproc, align=left, minimum width=\smbwd, minimum height=1cm, text centered, draw=black, fill=orange!30]
\tikzstyle{arrow} = [thick,->,>=stealth]
\def\smbwd{2cm}

\journal{Computer Physics Communications}

\begin{document}

\begin{frontmatter}


\title{FORTRESS II: FORTRAN programs for solving coupled 
       Gross-Pitaevskii equations for
	spin-orbit coupled spin-2 Bose-Einstein condensate}

\author[label1]{Paramjeet Banger}
\author[label1]{Pardeep Kaur}
\author[label2]{Arko Roy}
\author[label1]{Sandeep Gautam\corref{author}}
\cortext[author] {Corresponding author.\\\textit{E-mail address:} 
2018phz0003@iitrpr.ac.in, 2018phz0004@iitrpr.ac.in, arko.roy@unitn.it,
sandeep@iitrpr.ac.in}
\address[label1]{Department of Physics, 
	Indian Institute of Technology Ropar, Rupnagar, Punjab 140001, India}
\address[label2]{INO-CNR BEC Center and Dipartimento di Fisica, 
    Universit{\`a} di Trento, 38123 Trento, Italy}

\begin{abstract}
We provide here a set of three OpenMP parallelized FORTRAN 90/95 programs to compute the 
ground states and the dynamics of trapped spin-2 Bose-Einstein condensates (BECs) with 
anisotropic spin-orbit (SO) coupling by solving a set of five coupled Gross-Pitaevskii 
equations using a time-splitting Fourier spectral method. Depending on the
nature of the problem, without any loss of generality, we have employed the Cartesian grid 
spanning either three-, two-, or one-dimensional space for numerical discretization. To 
illustrate the veracity of the package, wherever feasible, we have compared the numerical 
ground state solutions of the full mean-field model with those from the simplified scalar models.
The two set of results show excellent agreement, in particular, through the equilibrium density 
profiles, energies and chemical potentials of the ground-states. We have also 
presented  test results for OpenMP performance parameters like speedup and the efficiency of the 
three codes.
\end{abstract}



\end{frontmatter}

{\bf PROGRAM SUMMARY}

\begin{small}
\noindent
	{\em Program Title: FORTRESS II}                                           \\
	{\em Licensing provisions:} MIT\\
	{\em Programming language:} (OpenMP) FORTRAN 90/95\\
{\em Computer:} Intel\textsuperscript{\textregistered} Xeon\textsuperscript{\textregistered}
Platinum 8160 CPU @ 2.10GHz\\                            
{\em Operating system}: General\\
{\em RAM}: Will depend on array sizes.\\
{\em Number of processors used}: 8 processors for 1D code and 
16 processors for 2D and 3D codes\\
{\em External routines/libraries:} FFTW 3.3.8 and Intel\textsuperscript{\textregistered} Math Kernel Library. The later is optional but gives better performance. \\
{\em Journal reference of previous version:} None                  \\
{\em Nature of problem:} To solve the coupled Gross-Pitaevskii equations for 
a spin-2 BEC with an anisotropic spin-orbit coupling.\\
	{\em Solution method:}
We use the time-splitting Fourier spectral method to split the coupled
Gross-Pitaevskii equations into four sets of sub-equations. The resulting sub-equations 
are evolved in imaginary time to obtain the ground state of the system or in realtime
to study the dynamics.\\
\end{small}


\section{Introduction}
The advent of optical traps in cold-atom experiments in the last couple of decades has made 
it possible to investigate the spinor Bose-Einstein condensates (BECs) in finer detail. A spinor
BEC, a Bose-Einstein condensate with internal spin degrees of freedom was first realized with 
$f=1$ $^{23}$Na atoms confined in an optical trap \cite{stamper1998optical}, where $f$ is the 
total spin per atom. Later, the different ground-state phases 
\cite{spin2-phases-ciobanu,spin2-phases-ueda} and spin 
dynamics of $f=2$ ${^{87}}$Rb 
\cite{chang2004observation,schmaljohann2004dynamics,kuwamoto2004magnetic,widera2005coherent}
spinor BEC were also examined. In an optical trap, internal spin degrees of freedom of an atom 
representing $2f+1$ hyperfine sublevels corresponding to the spin projection quantum numbers 
$m_f = -f,-f+1,\ldots,+f$ are simultaneously trapped, which is an impossibility in magnetic traps.
The interplay of the inter-atomic interactions and the Zeeman terms leads to a rich equilibrium
phase-diagram for a spin-2 BEC \cite{Kawaguchi-phases}. 
Another feature of the spinor BECs distinguishing these
from the scalar BECs is the spin-mixing dynamics \cite{widera2006precision}. 
One of the most crucial development in
the last decade in the field of spinor BECs has been the experimental realization of spin-orbit
(SO) coupling \cite{lin2011spin}, thus paving the way for several novel studies in the field 
of spin-2 BECs \cite{Xu,Kawakami}. The theoretical proposals to realise SO coupling in spin-2 
BECs have also been put forward \cite{soc-proposals}.

 In the absence of thermal and quantum fluctuations at $T=0$ K, the mean-field approximation 
allows one to describe an $f=2$ spinor BEC by a set of five-coupled nonlinear Gross-Pitaevskii 
equations (CGPEs) \cite{kawaguchi2012spinor}. In general this coupled set of equations, termed as
the mean-field model, is not analytically solvable without resorting to approximations. Thus there
is a need for an efficient, robust, and flexible numerical tool which will aid the aforementioned 
studies. In this context, a wide range of numerical techniques have been employed in literature to
study spinor BECs \cite{spin1, spin2-wang11,spin2-wang14,spin-1/2-soc}. 
In our earlier work \cite{spin1-soc}, we developed a set 
of F90/95 codes to solve the mean-field model of SO-coupled $f=1$ spinor BEC with Rashba 
\cite{bychkov1984oscillatory}
SO coupling using time-splitting Fourier spectral method. In the present work, we discuss 
the Fourier-spectral method to solve the CGPEs for an SO-coupled spin-2 BEC, where the parts 
of Hamiltonian corresponding to spin-exchange
collisions and SO coupling have to be numerically dealt with using a different numerical approach 
vis-{\`a}-vis a spin-1 BEC. To briefly summarize the method, we first split the CGPEs into four
sub-sets of equations, where each set consists of five equations, using  Lie  operator splitting.
The method then involves solving aforementioned four sets of equations one after the other over 
the same time interval with solution to each set serving as the input to the following set of 
equations. In the absence of SO coupling, we also develop two- and three-component scalar models
which can be used to study the static properties of the system. These scalar models also serve
an important purpose of validating the results obtained with the present set of codes where the 
SO coupling can be switched on/off by the user. 

The program package consists of a set of three OpenMP parallelized
FORTRAN 90/95 programs. This can be used to either  (a) calculate stationary state solutions or 
(b) study dynamics of  homogeneous or trapped SO-coupled spin-2 BECs, in three-dimensional (3D),
quasi-two-dimensional (q2D), and quasi-one-dimensional (q1D) configurations. The two different
objectives are accomplished by evolving the CGPEs either in imaginary or real time, respectively.
We have provided the users the option of switching between these two {\em modes} within the codes.

The paper is organised as follows. In Sec. \ref{Mean-Field-Model}, we introduce the mean-field
model of an SO-coupled spin-2 BEC. The scalar models which can be used to study a spin-2 BEC in
the absence of SO coupling are described in Sec. \ref{Scalar-Models}. We describe the 
time-splitting spectral method to solve the CGPEs in Sec. \ref{Numerical-Method}, 
followed by the description of the three codes in Sec. \ref{Details-of-Programs}. 
We present the test 
results for OpenMP performance in Sec. \ref{Parallelization-Tests} and the numerical results in 
Sec. \ref{Numerical-Results}.  
\section{Coupled Gross-Pitaevskii equations for an SO-coupled spin-2 BEC-Mean-field model}
\label{Mean-Field-Model}
The quantum and thermal fluctuations in an SO-coupled spin-2 BEC at $T=0$K can be neglected,
and the system is very well described by the following set of coupled Gross-Pitaevskii equations 
(CGPEs) in dimensionless form \cite{kawaguchi2012spinor,bychkov1984oscillatory}
\begin{subequations}
\begin{eqnarray}
i\frac{\partial \phi_{\pm 2}({\bf x},t)}{\partial t} &=& \mathcal{H} \phi_{\pm 2}({\bf x},t) + 
\tau_0 {\rho}({\bf x},t)  \phi_{\pm 2}({\bf x},t) + \tau_1 \{F_{\mp}({\bf x},t) \phi_{\pm 1}({\bf x},t) 
\pm 2 F_{z}({\bf x},t) \phi_{\pm 
2}({\bf x},t)\}\nonumber\\  
&&+\tau_2 \frac{\Theta({\bf x},t) \phi_{\mp 2}^*({\bf x},t)}{\sqrt{5}}+ \Gamma_{\pm 2}({\bf x},t), 
\label{cgpet3d-1}\\
i\frac{\partial \phi_{\pm 1}({\bf x},t)}{\partial t} &=& \mathcal{H} \phi_{\pm 1}({\bf x},t) 
+ \tau_0 {\rho}({\bf x},t)\phi_{\pm 1}({\bf x},t) + \tau_1 \left(\sqrt{\frac{3}{2}} F_{\mp}({\bf x},t)
\phi_{0}({\bf x},t) +F_{\pm}({\bf x},t) \phi_{\pm 2}({\bf x},t)\right.\nonumber\\ 
&&\left.\pm F_{z}({\bf x},t) \phi_{\pm 1}({\bf x},t) \right)
- \tau_2 \frac{\Theta({\bf x},t) \phi_{\mp 1}^*({\bf x},t)}{\sqrt{5}} + \Gamma_{\pm 1}({\bf x},t), 
\label{cgpet3d-2}\\
i\frac{\partial \phi_0({\bf x},t)}{\partial t} &=& \mathcal{H} \phi_0({\bf x},t) 
+ \tau_0 {\rho}({\bf x},t)
\phi_0({\bf x},t) + \tau_1 {\sqrt{\frac{3}{2}}}\{F_{-}({\bf x},t) \phi_{-1}({\bf x},t) +  
F_{+}({\bf x},t) \phi_1({\bf x},t)\}\nonumber\\  
&&+ \tau_2 \frac{\Theta({\bf x},t) \phi_{0}^*({\bf x},t)}{\sqrt{5}}+\Gamma_0({\bf x},t), \label{cgpet3d-3}
\end{eqnarray}
\end{subequations}
where, suppressing the explicit dependence of component wavefunctions $\phi_j$'s on ${\bf x}$ 
and $t$,
\begin{eqnarray*}
\mathcal{H}&=&-\frac{\mathbf{\nabla^2}}{2}+V(\mathbf{x})\quad \Theta = 
\frac{2\phi_2 \phi_{-2} - 2\phi_1\phi_{-1}+ \phi_0^2}{\sqrt{5}},
\quad F_z = \sum_{j=-2}^2 j|\phi_j|^2\\
F_- &=& F_+^* = 2\phi_{-2}^* \phi_{-1} + \sqrt{6}\phi_{-1}^* 
\phi_0 + \sqrt{6} \phi_0^* \phi_1 + 2 \phi_2 \phi_1^*, 
\end{eqnarray*}
and $\rho = \sum_{j=-2}^2 |\phi_j|^2$ is the total density. In 3D, $\bf x$, Laplacian, trapping
potential $V({\bf x})$, interaction parameters $(\tau_0,\tau_1,\tau_2)$, and SO-coupling terms $\Gamma$'s
are defined as
\begin{subequations}
\begin{eqnarray}
{\bf x} &\equiv& (x,y,z),\quad\mathbf{\nabla^2} = \left(\frac{\partial}{\partial x^2} + \frac{\partial}{\partial y^2}
+ \frac{\partial}{\partial z^2}\right),\quad V({\bf x}) = \frac{ \alpha_x^2x^2 
+ \alpha_y^2y^2 + \alpha_z^2z^2}{2}\\
\tau_0 &=& \frac{4\pi N(4 a_2 +3 a_4)}{7 a_{\rm osc}},\quad
\tau_1 = \frac{4\pi N( a_4 - a_2)}{7 a_{\rm osc}},\quad
\tau_2 = \frac{4\pi N( 7 a_0 - 10 a_2 +3a_4 )}{7 a_{\rm osc}},\\
\Gamma_{\pm 2}&=&-i\left(\gamma_x\frac{\partial\phi_{\pm 1}}
{\partial x}\mp i\gamma_y\frac{\partial\phi_{\pm 1}}{\partial y}\pm 2\gamma_z
\frac{\partial\phi_{\pm 2}}{\partial z}\right),\\
\Gamma_{\pm 1}&=&-i \left(  \gamma_x\frac{\partial\phi_{\pm 2}}{\partial x} 
+\sqrt{\frac{3}{2}}
\gamma_x\frac{\partial\phi_0}{\partial x} 
\pm i\gamma_y\frac{\partial\phi_{\pm 2}}
{\partial y} \mp i\sqrt{\frac{3}{2}}\gamma_y\frac{\partial\phi_0}{\partial y} 
\pm \gamma_z\frac{\partial\phi_{\pm 1}}{\partial z}\right)\\
\Gamma_{0}&=&-i\left(\sqrt{\frac{3}{2}}
\gamma_x\frac{\partial\phi_1}{\partial x} +\sqrt{\frac{3}{2}}\gamma_x
\frac{\partial\phi_{-1}}{\partial x} 
+ i{\sqrt{\frac{3}{2}}}\gamma_y\frac{\partial
\phi_1}{\partial y} -i\sqrt{\frac{3}{2}}\gamma_y\frac{\partial\phi_{-1}}
{\partial y}\right),
\end{eqnarray}
\end{subequations}
where $\alpha_\nu$ and $\gamma_\nu$ with $\nu =x,y,z$ are the anisotropy 
parameters of trapping potential and SO coupling, respectively;
$N$ is the total number of atoms; and $a_0, a_2, a_4$ are the $s$-wave scattering
lengths in total spin 0, 2 and 4 channels, respectively.

When a spin-2 BEC is strongly confined along one direction, say $z$, as compared
to other two, i.e. $\omega_z\gg \omega_z\sim \omega_y$, then one can approximate
Eqs. (\ref{cgpet3d-1})-(\ref{cgpet3d-3}) by quasi-two-dimensional (q2D) equations
which can obtained by substituting
\begin{subequations}
\begin{eqnarray}
{\bf x} &\equiv& x,y,\quad\mathbf{\nabla^2} = \left(\frac{\partial}{\partial x^2} + \frac{\partial}{\partial y^2}
\right),\quad V({\bf x}) = \frac{ \alpha_x^2x^2 
+ \alpha_y^2y^2}{2}\\
 \tau_0 &=& \sqrt{\frac{{\alpha_z}}{2\pi}} \frac{4\pi N(4 a_2 +3 a_4)}
{7 a_{\rm osc}},\quad
 \tau_1 = \sqrt{\frac{{\alpha_z}}{2\pi}}\frac{4\pi N(a_4-a_2)}
{7 a_{\rm osc}},\quad   
\tau_2 = \sqrt{\frac{{\alpha_z}}{2\pi}}\frac{4\pi N(7a_0-10a_2+3a_4 )}
{7 a_{\rm osc}}\\
\Gamma_{\pm 2}&=&-i\left(\gamma_x\frac{\partial\phi_{\pm 1}}
{\partial x}\mp i\gamma_y\frac{\partial\phi_{\pm 1}}{\partial y}\right),\\
\Gamma_{\pm 1}&=&-i \left(  \gamma_x\frac{\partial\phi_{\pm 2}}{\partial x} 
+\sqrt{\frac{3}{2}}
\gamma_x\frac{\partial\phi_0}{\partial x} 
\pm i\gamma_y\frac{\partial\phi_{\pm 2}}
{\partial y} \mp i\sqrt{\frac{3}{2}}\gamma_y\frac{\partial\phi_0}{\partial y} 
\right)\\
\Gamma_{0}&=&-i\left(\sqrt{\frac{3}{2}}
\gamma_x\frac{\partial\phi_1}{\partial x} +\sqrt{\frac{3}{2}}\gamma_x
\frac{\partial\phi_{-1}}{\partial x} 
+ i{\sqrt{\frac{3}{2}}}\gamma_y\frac{\partial
\phi_1}{\partial y} -i\sqrt{\frac{3}{2}}\gamma_y\frac{\partial\phi_{-1}}
{\partial y}\right).
\end{eqnarray}
\end{subequations}
Similarly, if the BEC is strongly confined along two directions, say $y$ and $z$, as compared
to third one, i.e. $\omega_y\sim \omega_z \gg \omega_x$, then one can approximate
Eqs. (\ref{cgpet3d-1})-(\ref{cgpet3d-3}) by quasi-one-dimensional (q1D) equations
which can obtained by substituting
\begin{subequations}
\begin{eqnarray}
{\bf x} &\equiv& x,\quad \mathbf{\nabla^2} = \frac{\partial}{\partial x^2},
\quad V({\bf x}) = \frac{ \alpha_x^2x^2}{2}\\
\tau_0 &=& \sqrt{\alpha_y \alpha_z}\frac{2 N(4 a_2 +3 a_4)}
{7 a_{\rm osc}},\quad
\tau_1 =\sqrt{\alpha_y \alpha_z}\frac{2 N(a_4-a_2)}
{7 a_{\rm osc}},\quad
\tau_2=\sqrt{\alpha_y \alpha_z}\frac{2 N(7a_0-10a_2+3a_4 )}
{7 a_{\rm osc}}\\
\Gamma_{\pm 2}&=&-i\left(\gamma_x\frac{\partial\phi_{\pm 1}}
{\partial x}\right),\\
\Gamma_{\pm 1}&=&-i \left(  \gamma_x\frac{\partial\phi_{\pm 2}}{\partial x} 
+\sqrt{\frac{3}{2}}
\gamma_x\frac{\partial\phi_0}{\partial x} 
\right)\\
\Gamma_{0}&=&-i\left(\sqrt{\frac{3}{2}}
\gamma_x\frac{\partial\phi_1}{\partial x} +\sqrt{\frac{3}{2}}\gamma_x
\frac{\partial\phi_{-1}}{\partial x}\right).
\end{eqnarray}
\end{subequations}

The energy of the SO-coupled spin-2 BEC is given as
\label{energyfun}
\begin{eqnarray}
E &=& \int d\textbf{x} 
\left[\left\{\sum_{j=-2}^{+2}\phi_j^{*}\left(-\frac{1}{2}\nabla^2 + V\right)\phi_j\right\}
+ \frac{\tau_0}{2}{\rho}^2  + \frac{\tau_1}{2}|{\bf F}|^2 + 
\frac{\tau_2}{2}|\Theta|^2 +\sum_{j=-2}^{+2}\phi_{j}^*\Gamma_{j} \right],
\end{eqnarray}
where $|{\bf F}|^2 = F_{+}F_{-} + F_z^2$. The energy along with norm 
${\cal N} = \int \rho d{\bf x}$ are two conserved quantities of an SO-coupled spin-2 BEC. The 
dimensionless formulation of the mean-field model, i.e. 
Eqs. (\ref{cgpet3d-1})-(\ref{cgpet3d-3}), ensures that 
${\cal N}$ is set to unity. In the absence of SO coupling, one more quantity longitudinal 
magnetization ${\cal M} = \int F_z d{\bf x} $ is also conserved. The time-independent variant
of Eqs. (\ref{cgpet3d-1})-(\ref{cgpet3d-3}) can be obtained by substituting 
$\phi_j({\bf x},t) = \phi_j({\bf x})e^{-i\mu_j t}$, where $\mu_j$'s are the chemical potentials
of the individual components. The conservation (non-conservation) of magnetization in the absence
(presence) of SO coupling is elaborated in Appendix.

\section{Scalar models for spin-2 BEC in the absence of SO coupling}
\label{Scalar-Models}
\subsection{Scalar model for ferromagnetic spin-2 BEC}
In the ferromagnetic domain, $\tau_1 <0$ and $\tau_2>20\tau_1$, a spin-2 BEC in the absence
of SO coupling has the component wavefunctions which are the multiples of 
a single wave function for the ground state \cite{gautam2015analytic}, i.e.
\begin{equation}
\phi_j({\bf x},t) = \beta_j\phi_{\rm DM}({\bf x},t) = |\beta_j|e^{i(\theta_j+\mu_j t)}
         \phi_{\rm DM}({\bf x}) ,\quad j = \pm2, \pm1,0 \label{DM},
\end{equation}
where $\beta_j$'s in general are complex.  
The $\beta$'s can be calculated by minimizing the $\tau_1$ and $\tau_2$ dependent 
energy terms under the constraints of fixed ${\cal N}$ and ${\cal M}$ 
and are \cite{gautam2015analytic}
\begin{equation}
|\beta_{\pm2}| = \frac{\left(2\pm{\cal M}\right)^2}{16},\quad
|\beta_{\pm1}| = \frac{\sqrt{4 -{\cal M}^2}(2\pm {\cal M})}{8},
\quad
|\beta_{0}| = {\frac{1}{8}}\sqrt{\frac{3}{2}}(4 - {\cal M}^2)
\label{spin2_beta},
\end{equation}
provided
\begin{equation}
\theta_2+\theta_{-2} - \theta_1-\theta_{-1} = 2p \pi,\quad
2\theta_0- \theta_1-\theta_{-1} = 2q \pi,\quad
\theta_0- 2\theta_1+\theta_2 = 2r \pi,\label{pr}
\end{equation}
where $p,q,r$ are integers.
Using Eqs. (\ref{DM}), (\ref{spin2_beta}), and 
(\ref{pr}) in Eqs. (\ref{cgpet3d-1})-(\ref{cgpet3d-3}) leads to 
the decoupling of the five CGPEs into five identical decoupled equations or one unique equation
known as single decoupled mode equation given as \cite{gautam2015analytic}
\begin{equation}
\mu  \phi_{\rm DM}({\bf x}) =\left(\frac {-\nabla ^2}{2} + V({\bf x})+ 
g|\phi_{\rm DM}({\bf x})|^2 \right)\phi_{\rm DM}({\bf x}), \label{SM1}
\end{equation}
where $g = \tau_0 + 4\tau_1$, and $\phi_{\rm DM}$ is the decoupled mode (DM) wavefunction. 
Thus, a ferromagnetic BEC in the absence of spin-orbit coupling can also be described by a 
single component scalar BEC model, and in the present work we use this model to validate 
the results from the full mean-field model described by Eqs. (\ref{cgpet3d-1})-(\ref{cgpet3d-3}). 
In the rest of the manuscript, we term Eq. (\ref{SM1}) as the single-component scalar model 
(SCSM).

\subsection{Scalar model for antiferromagnetic and cyclic spin-2 BECs}
For an antiferromagnetic system, $\tau_2<0$ and $\tau_2<20\tau_1$, the energy minimization 
corresponds to
minimization of $\tau_2$ dependent energy terms \cite{gautam2015analytic}. Assuming the system to be
uniform with a fixed particle density $\rho$, the minimization of $\tau_2$ dependent energy terms
under the constraints of fixed 
${\cal N}$ and ${\cal M}$ leads to \cite{gautam2015analytic}
\begin{subequations}
\begin{eqnarray}
\int \rho_{\pm 2} d{\bf x} &=& \frac{2\pm{\cal M}}{4},\label{beta2_af}\\
\phi_{\pm 1}({\bf x}) &=& \phi_0({\bf x}) = 0\label{beta1_af}.
\end{eqnarray}
\end{subequations}
Using Eq. (\ref{beta1_af}) in time-independent variant of Eqs. 
(\ref{cgpet3d-1})-(\ref{cgpet3d-3}) leads to 
the two coupled GP equations \cite{gautam2015analytic},
\begin{subequations}\label{SM2}
\begin{eqnarray}
\mu_2\phi_2 &=& \left[\mathcal{H} + (\tau_0 + 4 \tau_1) |\phi_2|^2 + 
\left(\tau_0 - 4\tau_1 + \frac{2}{5}\tau_2\right)|\phi_{-2}|^2\right]\phi_2,\label{gp1}\\
\mu_{-2}\phi_{-2} &=& \left[\mathcal{H} + (\tau_0 + 4 \tau_1) |\phi_{-2}|^2 + 
\left(\tau_0 - 4\tau_1 + \frac{2}{5}\tau_2\right)|\phi_{2}|^2\right]\phi_{-2}\label{gp2}.
\end{eqnarray}
\end{subequations}
Thus, the ground state of an antiferromagnetic BEC in the absence of spin-orbit coupling may 
also be described by a two-component scalar BEC model. In the rest of the manuscript, we term 
Eqs. (\ref{beta2_af})-(\ref{beta1_af}) and (\ref{SM2}) as the two-component scalar model 
(TCSM) for an antiferromagnetic system.

Similarly, in the cyclic phase, $\tau_1>0$ and $\tau_2>0$, and for energy minimization one
needs to minimize both $\tau_1$ and $\tau_2$ dependent energy terms.
Using the uniform system approximation, the minimization of $\tau_1$ and 
$\tau_2$ dependent energy terms under the constraints of fixed
${\cal N}$ and ${\cal M}$ leads to two degenerate states for all possible
magnetizations \cite{gautam2015analytic}. First of these states has
\begin{subequations}
\begin{eqnarray}
\int \rho_{2} d{\bf x} &=& \frac{1+{\cal M}}{3},\quad \int \rho_{- 1}d{\bf x} =  \frac{2-{\cal M}}{3}\label{betam1_c},\\
\phi_{1}({\bf x}) &=& \phi_{0}({\bf x}) ~=~\phi_{-2}({\bf x})~=~ 0\label{beta1_c}, 
\end{eqnarray}
\end{subequations}
and the second has
\begin{subequations}
\begin{eqnarray}
\int \rho_{\pm 2} d{\bf x} &=& \left(\frac{2 \pm {\cal M}}{4}\right)^2,\quad
\int \rho_0d{\bf x} =  \frac{4-{\cal M}^2}{8},\\
\phi_{1}({\bf x}) &=& \phi_{-1}({\bf x})~=~ 0. 
\end{eqnarray}
\end{subequations}
The latter of these states would lead to the three component model and is not
considered in the present work. Using Eq. (\ref{beta1_c}) in time-independent variant of Eqs.
(\ref{cgpet3d-1})-(\ref{cgpet3d-3}), one again
obtains two coupled GP equations \cite{gautam2015analytic}
\begin{subequations}\label{SM3}
\begin{eqnarray}
\mu_2\phi_2 &=& \left[\mathcal{H} + (\tau_0 + 4 \tau_1) |\phi_2|^2 + 
\left(\tau_0 - 2\tau_1 \right)|\phi_{-1}|^2\right]\phi_2,\label{gp1_c}\\
\mu_{-1}\phi_{-1} &=& \left[\mathcal{H} + (\tau_0 + \tau_1) |\phi_{-1}|^2 + 
\left(\tau_0 - 2\tau_1 \right)|\phi_{2}|^2\right]\phi_{-1}\label{gp2_3}.
\end{eqnarray}
\end{subequations}
Eqs. (\ref{betam1_c})-(\ref{beta1_c}) and (\ref{SM3}), constitute the TCSM 
for a cyclic BEC. The scalar models are much easier to solve as the Hamiltonian
consists of only diagonal terms \cite{Adhikari,chang2005gauss}. 

\section{Numerical method: Time-splitting Fourier Spectral method}
\label{Numerical-Method}
We use the time-splitting Fourier spectral method to solve the CGPEs \cite{spin1-soc}. 
Here, we elaborate the method to solve Eqs. (\ref{cgpet3d-1})-(\ref{cgpet3d-3}) for q1D case
as an  archetypal system. The extension to q2D and 3D is straightforward.
The CGPEs (\ref{cgpet3d-1})-(\ref{cgpet3d-3}) can be written in matrix form
as
\begin{equation}
i\frac{\partial\Phi}{\partial t} =( H_{\rm SP} + H_{\rm SE}+ H_{\rm SOC}+H_{\rm KE} )\Phi, \label{split}
\end{equation} 
where $H_{\rm KE}$ is a diagonal matrix consisting of kinetic energy operators, $H_{\rm SOC}$ is
matrix operator corresponding to spin-orbit coupling, $H_{\rm SE}$ consists of off-diagonal 
interaction terms, and $H_{\rm SP}$ is a diagonal matrix consisting of trapping 
potential plus diagonal interaction terms. 
These $5\times5$ matrices are defined as
\begin{subequations}
\begin{align}
H_{\rm KE} &= \text{diag}\left(-\frac{\partial^2_x}{2},-\frac{\partial^2_x}{2},-\frac{\partial^2_x
}{2},-\frac{\partial^2_x}{2},-\frac{\partial^2_x}{2}\right),\quad H_{\rm SP}=\text{diag}\left( 
H_{+ 2},H_{+1},H_0,H_{-1},H_{-2}\right);\\
H_{\rm SOC} &= \begin{pmatrix}
0 & \partial_{x} & 0 & 0 & 0 \\
\partial_{x} & 0  & \sqrt{\frac{3}{2}}\partial_{x} & 0 & 0\\
0  & \sqrt{\frac{3}{2}}\partial_{x} &  0  & \sqrt{\frac{3}{2}}\partial_{x} & 0\\
0 & 0  & \sqrt{\frac{3}{2}}\partial_{x} & 0 & \partial_{x}\\
0 & 0 & 0 & \partial_{x} & 0. 
\end{pmatrix};~
H_{\rm SE} = 
\begin{pmatrix}
0 & H_{12} &H_{13}  & 0 &0\\
H_{12}^* &0 &H_{23}  & 0 &0\\
H_{13}^* & H_{23}^* & 0 & H_{34} &H_{35} \\
0 &0  & H_{34}^* &0 &H_{45} \\
0 &0  &H_{35}^* & H_{45}^*  & 0
\end{pmatrix}{\label{SEmatrix}};
\end{align}
\end{subequations}
where\quad $H_{\pm 2} = V/2  + \tau_0 {\rho} \pm 2\tau_1 F_{z} + (2/5) \tau_2 
|\phi_{\mp2}|^2$, $H_{0} = V/2  + \tau_0 {\rho} + (1/5)\tau_2 |\phi_{0}|^2$, 
$H_{\pm 1} = V/2  + \tau_0 {\rho} \pm \tau_1 F_{z} + (2/5) \tau_2 |\phi_{\mp1}|^2$, and
$ H_{12} =  \tau_1 F_{-} - (2/5)\tau_2\phi_{-1} \phi_{-2}^*$,  
$ H_{13} = (1/5)\tau_2\phi_{0} \phi_{-2}^*$,  
$ H_{23} =  (\sqrt{6}/2)\tau_1 F_{-}- (1/5)\tau_2\phi_{0} \phi_{-1}^*$,
$ H_{34} = (\sqrt{6}/2)\tau_1 F_{-}- (1/5)\tau_2\phi_{1} \phi_{0}^*$,  
$ H_{35} = (1/5)\tau_2\phi_{2} \phi_{0}^*$, 
$H_{45} =  \tau_1 F_{-} - (2/5) \tau_2\phi_{2} \phi_{1}^*$.
Eq. (\ref{split}) is split into four equations by using the standard
Lie splitting. The numerical methods to solve equations corresponding to diagonal 
operators, $H_{\rm KE}$ and $H_{\rm SP}$, are discussed in detail in
Refs. \cite{spin1-soc,ravisankar2020spin}. Hence, we focus on the solving the (split) equations corresponding
to off-diagonal operators. To solve the equation corresponding to $H_{\rm SOC}$, we first
take the Fourier transform of the equation to obtain
\begin{equation}
i\frac{\partial {\hat \Phi}(k_x,t)}{\partial t} = \hat{H}_{\rm SOC}\hat{\Phi}(k_x,t),\label{SOCpartf}
\end{equation}
where $\hat H_{\rm SOC}$ can be obtained from $H_{\rm SOC}$ by substituting 
$\partial_x $ by $k_x$
The formal solution of Eq. (\ref{SOCpartf}) is
\begin{eqnarray}
{\hat \Phi}(k_x,t +\delta t) &=& \exp(-i\delta t \hat{H}_{\rm SOC})\hat{\Phi}(k_x,t),\nonumber\\
                             &=& \exp(-i\delta t \hat{P} \hat{D} \hat{P}^{-1})\hat{\Phi}(k_x,t),\nonumber\\
                             &=&\hat{P}  \exp(-i\delta t \hat{D} ) \hat{P}^{-1} \hat{\Phi}(k_x,t),
\end{eqnarray}
where $D$ is the diagonal matrix.  The solution of split equation for $H_{\rm SE}$ is approached
similarly with one difference that $H_{\rm SE}$ is time-dependent. 
Taking this into account, solution to equation for $H_{\rm SE}$ is 
\cite{spin2-wang11}

\begin{eqnarray}
\Phi(x,t +\delta t) &=& \exp\left(-i\int_t^{t+\delta t} H_{\rm SE}(x,t)dt\right)\Phi(x,t),\nonumber\\
                             &\approx& \exp\left[-i{\delta t} \frac{\left\{H_{\rm SE}(x,t) 
                             + H_{\rm SE}^{fE}(x,t+\delta t)\right\}}{2}\right]\Phi(x,t),\nonumber\\
                             &=& \exp(-i\delta t S A S^{-1})\Phi(x,t),\nonumber\\
                             &=&S  \exp(-i\delta t A ) S^{-1} \Phi(x,t),
\end{eqnarray}
where $ H_{\rm SE}^{fE}(x,t+\delta t)$ is estimated by the forward Euler ($fE$) method, and 
$A$ is the diagonal matrix.

\section{Details about the programs} 
\label{Details-of-Programs}
Here we describe the set of three codes written in FORTRAN 90/95 programming language to solve 
CGPEs (\ref{cgpet3d-1})-(\ref{cgpet3d-3}) as per the numerical method described in the previous
section. These three programs, namely  
\textbf{imretim1D\_spin2.f90}, \textbf{imretime2D\_spin2.f90}, and 
\textbf{imretime3D\_spin2.f90} correspond to q1D, q2D, and 3D systems, respectively. We employ
the  time-splitting Fourier spectral method to solve CGPEs, and the resultant set of equations
are evolved over imaginary or real time to study the stationary states or 
dynamics, respectively. Herein we introduce the prospective user of software package to the
various {\em modules, subroutines} and {\em functions}, and {\em input}/{\em output files}. 
We use {\em imretime1D\_spin2.f90} as a typical example to introduce these various 
constituents  of the code.

\subsection{Modules} 
The various input parameters needed by the main program are defined in the three modules at the
top of each program. These modules are BASIC\_DATA, CGPE\_DATA, and SOC\_DATA.
The parameters which the user may need to modify depending on his/her problem of interest
are defined in these three modules.

\subsubsection*{BASIC\_DATA}
The number of one-dimensional spatial grid points NX defined in this module has to be chosen 
consistent with the spatial-step DX so that the total spatial extent LX=NX $\times$ DX is 
sufficiently larger than the size of the condensate. The number of OpenMP and FFTW threads to be
used are defined by OPENMP\_THREADS and FFTW\_THREADS in this module. 
The integer parameter NITER, denoting the maximum number of time iterations should be chosen 
sufficiently large to obtain the requisite convergence in imaginary time propagation.
NSTP representing the number of iterations after which transient wavefunctions are written, and
STP representing the number of iterations after which energy, chemical potentials, and rms sizes
are calculated should be chosen by user as per the need of the problem.

\subsubsection*{CGPE\_DATA}
The scattering lengths (A0, A2, A4) in Bohr radius, mass of atom M in atomic mass unit 
corresponding to the atomic species of spin-2 BEC, total number of atoms NATOMS, trapping 
frequencies along three axes (NUX, NUY, NUZ) are the user defined variables in the module. 
In addition to these, an integer parameter SWITCH\_IM has to be set equal to 1 or 0 for 
imaginary- or real-time propagation, respectively. The component wavefunctions, 
PHI(1:NX, 1:5)$\equiv\phi_j(x)$ and their discrete Fourier transforms 
PHIF(1:NX, 1:5)$\equiv\hat{\phi}_j(k_x)$ are declared in this module.
In imaginary-time propagation mode, the execution of the program is stopped if the convergence 
criterion, $\max |\phi_j(x,t) - \phi_j(x,t-\delta t)|/(2 \delta t)$, falls below the user 
defined tolerance (TOL) which is set to $10^{-6}$ in the module. 
In the absence of SO coupling, OPTION\_FPC = 1, 2 or 3 allows the user to choose a suitable 
initial guess for ferromagnetic, polar or cyclic phases, respectively, whereas OPTION\_FPC can be
set to 4 to use Gaussian initial guess wavefunctions in the presence of SO coupling. 

\subsubsection*{SOC\_DATA}
The SWITCH\_SOC parameter in this module has to be set equal to 1 or 0 in the 
presence or absence of SO coupling, respectively. The full list and description of parameters or variables defined or declared in the above 
three modules are listed in Table \ref{module}.

\subsubsection*{FFTW\_DATA}
The variable types of the input and output arrays used in FFTW subroutines to calculate forward 
and backward discrete Fourier transform, the necessary FFTW plans, and (FFTW) thread 
initialization variable are declared in this module. The module uses the FFTW3 module from the 
FFTW software library which defines the various variable types needed by the FFTW subroutines 
\cite{fftw}. FFTW\_DATA is not required to be modified by the user.

\begin{table}[!hbtp]
\caption{Description of various modules}
\begin{center}
\begin{tabular}{ p{2cm} p{3cm}  p{8cm}  }
\hline
\multicolumn{1}{c}{Module name}&
\multicolumn{1}{l}{Parameter/Variable}&
\multicolumn{1}{l}{Description}\\
\hline
{BASIC\_DATA} &PI, CI& $\pi$ and $\sqrt{-1}$\\
& NITER & Total number of time iterations\\
& NSTP & Number of iterations after which component densities $\rho_j$
and their corresponding phases are written\\
&OPENMP\_THREADS& Number of OpenMP threads \\
&FFTW\_THREADS& Number of FFTW threads\\
&NX& Number of spatial-grid points in $x$-direction\\
&DX, DT &Spatial and temporal step-sizes\\
& LX& Spatial domain chosen to solve the CGPEs\\
& STP& Number of iterations after which energy, chemical potentials, and rms sizes corresponding 
to each component are calculated\\
&AMU, HBAR& Atomic mass unit and reduced Planck's constant \\
& CDT & Complex variable defined as -idt or dt in imaginary or real-time propagation,
respectively\\
\hline
 CGPE\_DATA &M, A0, A2, A4 &Mass of atom in kg and scattering lengths ($a_0,a_2,a_2$) in 
 meters corresponding to total spin channels 0, 2 and 4, respectively\\
 &NUX, NUY, NUZ & Trapping frequencies in Hz along $x, y,$ and $z$ axes, respectively\\
 &ALPHAX, ALPHAY, ALPHAZ & Anisotropy parameters ($\alpha_x,\alpha_y,\alpha_z$) \\
 &NATOMS & Total number of atoms\\
 &NTRIAL & Maximum number of iterations for Newton-Raphson method to solve Eqs. 
  (\ref{norm-6})-(\ref{norm-7})\\ 
 &X, X2 &  Real 1D arrays for spatial grid and its square\\
 &KX & Real 1D array for Fourier grid \\
 &V, R2 & Real 1D arrays for trapping potential and $r^2$\\
 &PHI, PHIF & Complex 2D arrays for wavefunctions in real and Fourier space\\
 &AOSC & Real variable for oscillator length \\
 &OMAGAM & Real variable for angular trapping frequency $\omega_x$ along $x$ axis 
        chosen to scale the frequencies\\
 &TAU & Real 1D array variable with three elements TAU(0), TAU(1), TAU(2) for three 
        interaction parameters\\
 &MAG & Real variable for magnetization\\
 &SWITCH\_IM& It is set to 1 or 0 to choose imaginary or real-time propagation\\
 &OPTION\_FPC& Option to choose the initial guess solution\\
 &N1, N2, N3, N4, N5& Real variables for component norms
 \\\hline
   SOC\_DATA  & SWITCH\_SOC & It is set to 1 or 0 for non-zero or zero SO coupling, respectively\\
    & GAMMAX & Strength of SO coupling along $x$ direction \\
\hline
\end{tabular}
\label{module}
\end{center}
\end{table}
\begin{figure}[!hbtp]
    \centering
    \includegraphics[width=12.5cm]{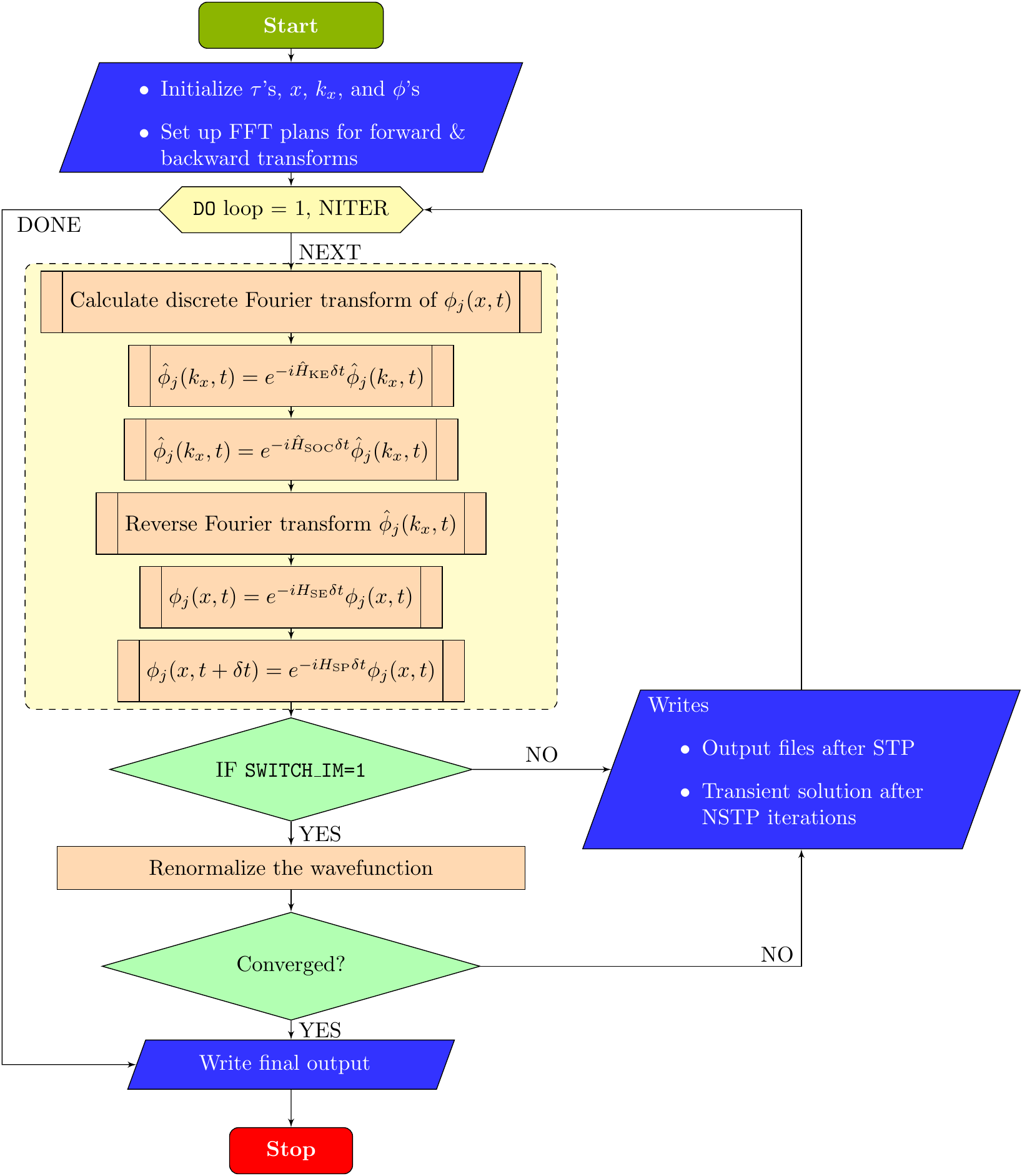}
    \caption{Flowchart illustrating the implementation of numerical procedure}
    \label{flowchart}
\end{figure}
\begin{table}[!htbp]
\caption{Description of various subroutines and functions involved in program and their usage.}
\begin{center}
\begin{tabular}{ p{3.0cm} p{1.5cm}  p{8cm}}
\hline
\multicolumn{1}{l }{Name}&
\multicolumn{1}{c}{Type}&
\multicolumn{1}{c}{Description}\\
\hline

INITIALIZE & Subroutine & Defines the spatial and Fourier grids, trapping potential, and
initializes the component wave functions\\
CREATE\_PLANS & Subroutine & Creates FFTW plans (with threads) for forward and backward 
transforms\\
DESTROY\_PLANS & Subroutine & Destroys FFTW plans for forward and backward transforms\\
FFT & Subroutine & Calculates the discrete forward Fourier transform 
\\
KE & Subroutine &  Solves the split-equation corresponding to $H_{\rm KE}$ in Fourier space\\
SOC & Subroutine & Solves the split-equation corresponding to $H_{\rm SOC}$ in Fourier space\\
BFT & Subroutine & Calculates the discrete backward Fourier transform of component wavefunctions\\
SE & Subroutine & Solves the split-equation corresponding to $H_{\rm SE}$\\
SP & Subroutine & Solves the split-equation corresponding to $H_{\rm SP}$\\
FXYZ & Subroutine & Calculates $F_x$ (FX), $F_y$ (FY) and $F_z$ (FZ) to evaluate 
$F_-$ (FMINUS) and $F_+$ (FPLUS)\\
MAT\_C & Subroutine & Calculates $H_{12}$ (C12), $H_{13}$ (C13), $H_{23}$ (C23), $H_{34}$ (C34), $H_{35}$ (C35) and $H_{45}$ (C45), i.e. the elements of $H_{SE}$.\\
NORMT & Subroutine &  Normalizes the total density to 1\\
NEWTON & Subroutine & Solves the non-linear Eqs. (\ref{norm-6})-(\ref{norm-7}) 
by using Newton Raphson method  \\
NORMC & Subroutine & Calculates the norm of individual components\\
RAD & Subroutine & Calculates the root mean square (rms) sizes of the five components\\
ENERGY & Subroutine & Calculates the five component chemical potentials 
$\mu$ (MU), energy $E$ (EN), and magnetization $\cal M$ (MZ)\\
SIMPSON & Function & Performs integration by Simpson's $1/3$ rule \\
DIFF &  Function &  Evaluates $df(x)/dx$ using nine 
point Richardson's extrapolation formula\\

\hline
\end{tabular}
\label{subroutine}
\end{center}
\end{table}
\subsection{Functions \& subroutines}
All the subroutines and functions, used in the three codes, and their specific tasks are listed 
in Table \ref{subroutine}. The user does not need to make any changes to these subroutines and 
functions. The overall organization of the various procedures for 'imagtime1D\_spin2.f90' is
illustrated in the flowchart provided in Fig.~\ref{flowchart}. The subroutine symbols in the 
flowchart, i.e. a rectangle with 
double-struck vertical edges, from the top to bottom, respectively, represent the successive 
calls to subroutines FFT, KE, SOC, BFT, SE, and SP.

The description of the various output files written by the codes is provided in Table 
\ref{OPfile}. Besides these output files, in realtime-propagation mode, user needs to provide 
an input file 'initial\_sol.dat' whose contents are in the same format as 
'solution\_file\_im.dat'.
\begin{table}[!htbp]
\caption{Description of various output files and their contents. '*' denotes 'im' or 're'.}
\begin{center}
\begin{tabular}{ p{3cm} p{2cm}  p{8cm}  }
\hline
\multicolumn{1}{l}{Name}&
\multicolumn{1}{l}{Time propagation}&
\multicolumn{1}{c}{Contents}\\
\hline
file1\_*.dat & imaginary/real & Various input parameters are written at the top of file. 
                Total norm, energy, chemical potentials and $|\phi_j|$'s at the origin are written after each NSTP iterations.\\
file2\_*.dat & imaginary/real & Time, energy and rms sizes for individual components after each
STP iterations\\ 
file3\_*.dat &imaginary/real & Time, norm of the individual components, sum of norms of individual components ,  and magnetization after 
each STP iterations \\
convergence.dat & imaginary & Number of iterations and convergence attained after each STP 
iterations. \\
tmp\_solution\_file.dat & imaginary/real & Component densities $\rho_j$ and their 
corresponding phases are written at every 
space point and updated after each NSTP iterations. \\
solution\_file\_*.dat & imaginary/real & Final component densities $\rho_j$ and their 
corresponding phases are written at every space point. \\\hline
\end{tabular}
\label{OPfile}
\end{center}
\end{table}
We have described here the 1D code, but the structure of code in 2D and 3D is identical.
The names and role of modules, subroutines, and functions in three codes are also same.
The main difference would be due to the fact that in 2D program, spatial grid consists of NX 
and NY points with spatial step sizes DX and DY along $x$ and $y$ directions, respectively. 
The resultant spatial domain along these directions would be LX = DX${\times}$NX and 
LY = DY${\times}$NY, respectively. Similarly, in 3D program, spatial grid consists of total
of NX$\times$NY$\times$NZ points with spatial step sizes of DX, DY, and DZ along three directions.

\section{OpenMP Parallelization}
\label{Parallelization-Tests}
\begin{figure}
\centering
\includegraphics[width=1.1\linewidth]{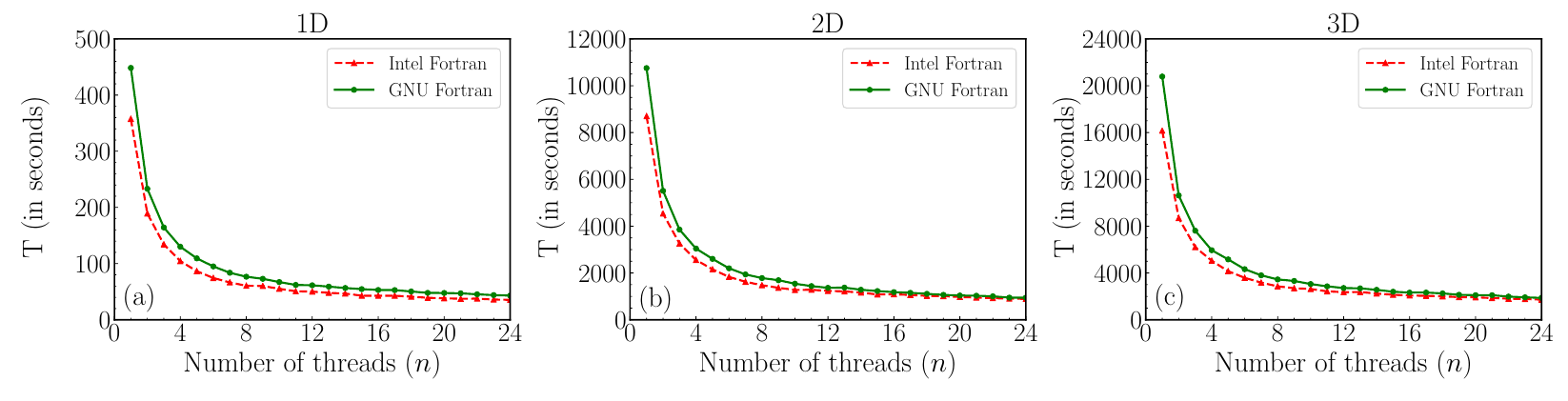}
\caption{(Color online) (a) Execution time for 1000 iterations (in
 seconds) as a function of number of threads for 1D code compiled with GNU Fortran 5.4.0 and Intel 
 Fortran 18.0.3 compilers for imaginary time propagation. (b) and (c) are the same for 2D 
 and 3D codes.}
\label{execution}
\end{figure}

\begin{figure}
    \includegraphics[width=1.15\linewidth]{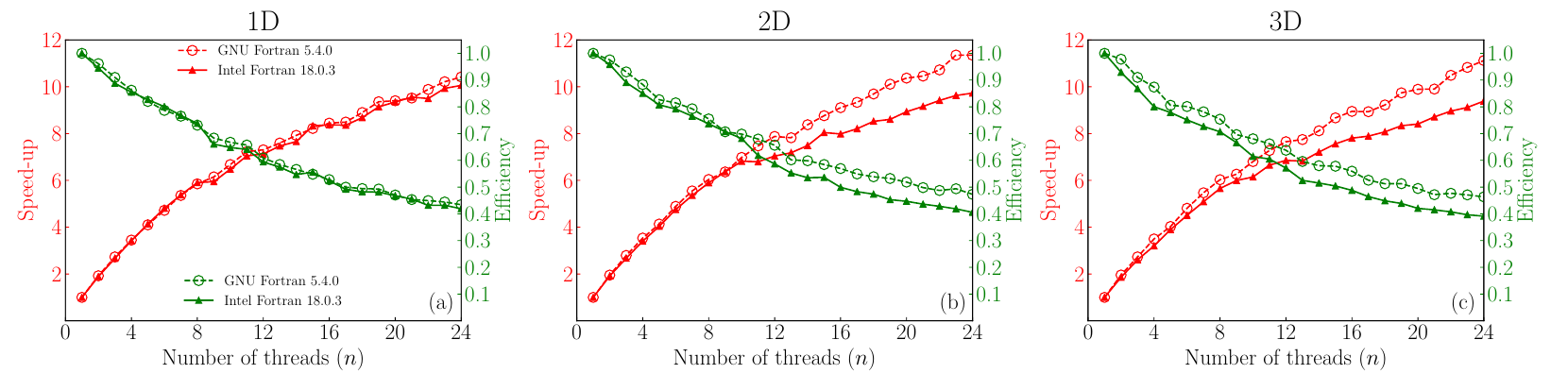}
    	\caption {(Color online) Speedup and efficiency as a function of number of threads (n) are shown for imaginary time propagation.  Figures (a)-(c) show the results for 1D, 2D and 3D codes, respectively.}
    \label{openmp}
\end{figure}

For the three codes, we have tested the performance of OpenMP parallelization for their imaginary
as well as real-time variants on a 24-core Intel\textsuperscript{\textregistered}
Xeon\textsuperscript{\textregistered} Platinum 8160 CPU@ 2.10 GHz processor. The OpenMP performances
of the imaginary time and real-time variants are quite similar. The array sizes considered to 
perform these parallelization tests are NX = 50000, NX = NY = 1024, and 
NX =  NY = NZ = 128 for 1D, 2D and 3D codes, respectively. We measured the elapsed wall clock 
time for 1000 iterations 
starting from the INITIALIZE subroutine and have not counted the time spent in opening/closing 
and reading/writing the data files. We have studied the performance of these codes with Intel 
Fortran 18.0.3 and GNU Fortran 5.4.0 compilers using upto 24 processors and confirmed that our 
OpenMP Fortran programs are optimized for both the compilers. A significant decrease in execution
time has been observed for all the codes as is quite clear from the test results presented for the 
imaginary time variants in Fig. \ref{execution}.
To get the quantitative estimate of OpenMP parallelization, we have calculated the speedup and
efficiency for all these codes for both the compilers, where speedup is defined as the ratio of 
execution time with 1 thread to the execution time with $n$ threads and efficiency is the ratio
of speedup to the number of threads. We have achieved an excellent speedup of above 
10 with 24 threads for 1D code and above 9 for  2D as well as 3D codes with aforementioned 
compilers as shown in Fig. \ref{openmp}. All these tests have been 
performed with non-zero value of SOC strength.

\section{Numerical Results}
\label{Numerical-Results}
In this section, we present the numerical results for energies, 
chemical potentials, and densities of the ground states of q1D, 
q2D and 3D spin-2 BECs. We also present the results for dynamics of a q1D BEC.
\subsection{Results for q1D systems}
\label{q1d}
Here we present the results for q1D BECs first in the absence of SO coupling and then in the 
presence of SO coupling followed by the results for dynamics in the presence of SO coupling.
\subsubsection*{Without SO coupling, $\gamma_x$  = 0}
\begin{figure}
 \centering
 \includegraphics[width=1\linewidth]{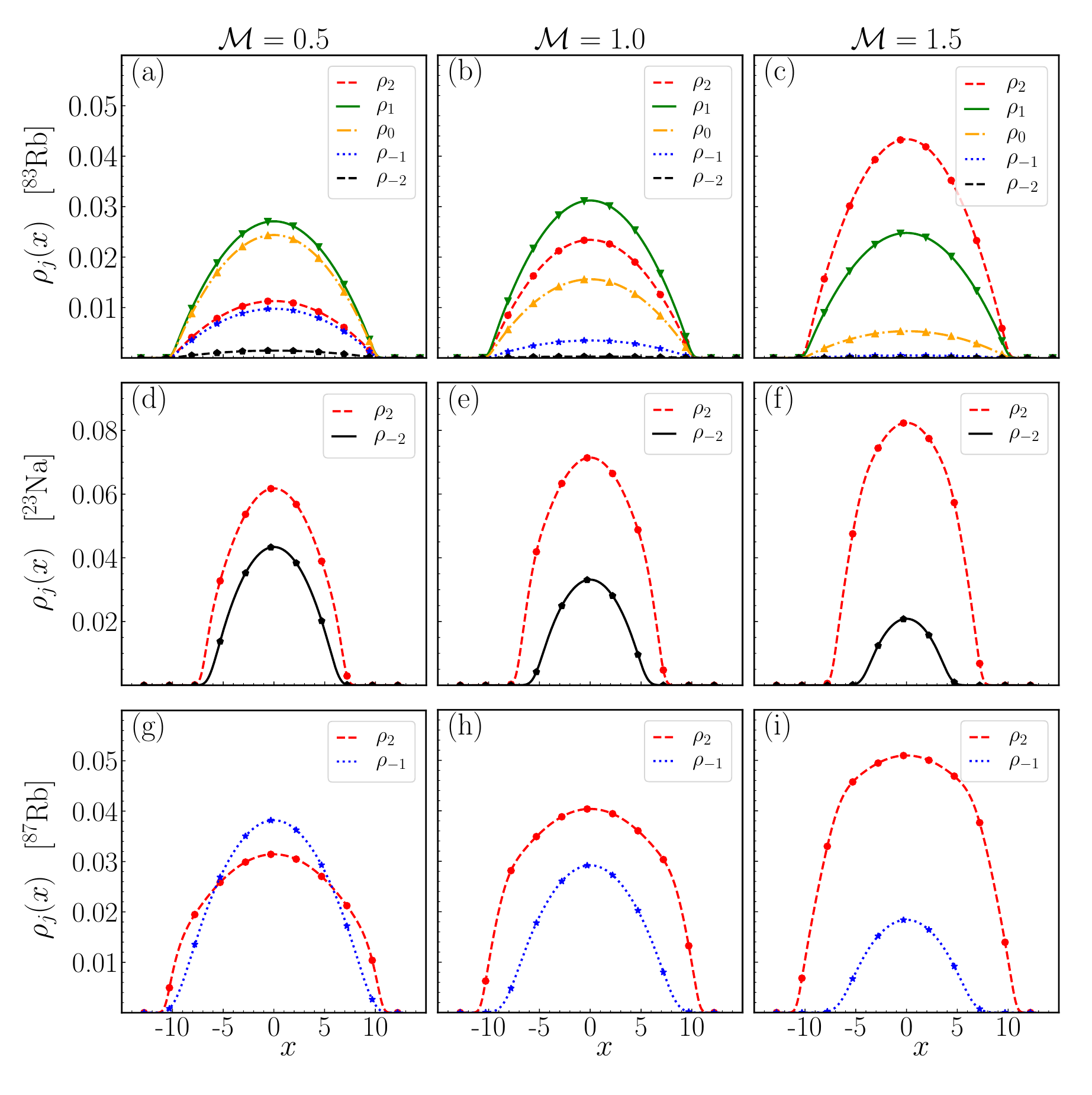}
 \caption {(Color online) (a)-(c) are numerically calculated ground state component densities 
 with the full mean-field model (indicated by different line styles) and the scalar model
 (indicated by different point styles) for a spin-2 BEC of $^{83}$Rb with different values of 
 ${\cal M}$. (d)-(f) and (g)-(i) are the same for $^{23}$Na and $^{87}$Rb, respectively.}
\label{fig1}
\end{figure}

We consider (a) $^{83}$Rb, (b) $^{23}$Na, and (c) $^{87}$Rb spin-2 BECs as 
the typical examples of ferromagnetic, anti-ferromagnetic and cyclic phases. 
The three scattering length values considered for these systems are 
\cite{spin2-phases-ciobanu,widera2006precision}
\begin{eqnarray*}
{\rm (a)}\quad a_0 &=& 83.0a_B,\quad a_2 = 82.0a_B,\quad a_4 = 81.0a_B;\\
{\rm (b)}\quad a_0 &=& 34.9a_B,\quad a_2 = 45.8a_B,\quad a_4 = 64.5a_B;\\
{\rm (c)}\quad a_0 &=& 87.93a_B,\quad a_2 = 91.28a_B,\quad a_4 = 99.18a_B,
\end{eqnarray*}
respectively. 
We consider $10000$ atoms of each of these systems trapped in q1D trapping potential
with $\omega_x = 2\pi\times 20$Hz, $\omega_y = \omega_z = 2\pi\times 400$Hz, and
thus $\alpha_x = 1$ and $\alpha_y = \alpha_z = 20$.
The oscillator lengths for the three systems are $2.47~\mu$m for (a), $4.69~\mu$m 
for (b), and $2.41~\mu$m for (c). The triplet of dimensionless interaction 
strengths $(\tau_0,\tau_1,\tau_2)$ are given as
\begin{eqnarray*}
{\rm (a)}\quad (\tau_0,\tau_1,\tau_2) &=& (699.62, -1.23, 4.90),\\
{\rm (b)}\quad (\tau_0, \tau_1,\tau_2) &=& (242.97, 12.06, -13.03),\\
{\rm (c)}\quad (\tau_0,\tau_1,\tau_2) &=& (831.26, 9.91, 0.31).
\label{param}
\end{eqnarray*}
For the conversion of dimensional variables to their dimensionless analogues,
we refer the reader to Ref. \cite{spin1-soc}.
For $^{83}$Rb the results obtained by solving Eqs. (\ref{cgpet3d-1})-(\ref{cgpet3d-3})
are compared with SCSM, viz. Eqs. (\ref{DM})-(\ref{SM1}), with 
$g = \tau_0 + 4\tau_1$.  
Similarly, for $^{23}$Na and $^{87}$Rb spin-2 BECs, the results are compared
with TCSM, viz. Eqs. (\ref{beta2_af})-(\ref{gp2}) for the former and 
Eqs. (\ref{betam1_c})-(\ref{beta1_c}) and Eq. (\ref{SM3}) for the latter.  The ground state 
chemical potentials and  energies obtained for q1D $^{83}$Rb  using the full mean-field and scalar
models are given in table \ref{table1} with various values of $\cal M$. Similarly, 
comparisons of chemical potentials and energies from both the models for q1D $^{23}$Na and $^{87}$Rb 
are presented in table \ref{table2} and \ref{table3}, respectively. The agreement between
the two set of results is excellent and is also evident from ground state density profiles for 
the three systems shown in Fig. \ref{fig1}.

\begin{table}[!h]
\caption{Ground state energies and chemical potential values of q1D $^{83}$Rb BEC obtained 
with spin-2 mean-field model and SCSM. The results have been obtained with $\Delta x = 0.05$ and
$\Delta t = 0.000125$ for different values of $\cal M$.}
\begin{center}
\begin{tabular}{| c | c | c | c | c | c |}
\hline
\multicolumn{1}{ |c| }{}& \multicolumn{1}{|c|}{Eqs. (\ref{cgpet3d-1})-(\ref{cgpet3d-3})}&
\multicolumn{1}{ |c| } {SCSM}&
\multicolumn{2}{|c|}{ $^{83}$Rb}\\
\hline
\multicolumn{1}{ |c| }{${\cal M}$}&
\multicolumn{1}{ |c| } {$\mu_{\pm 2}= \mu_{\pm 1} = \mu_0$}& 
\multicolumn{1}{ |c| }{$\mu=\mu_{\pm 2}= \mu_{\pm 1} = \mu_0$}&
\multicolumn{1}{ |c| } {$E$} &
\multicolumn{1}{ |c| }{$E$ - SCSM}\\
\hline
0-1.9& 51.3976 &51.3991 &30.8496&30.8496\\ \hline
\end{tabular}
\label{table1}
\end{center}
\end{table}
		
\begin{table}[!h]
\caption{
 Ground state energies and chemical potential values of q1D $^{23}$Na BEC obtained with 
 spin-2 mean-field model and TCSM. The results have been obtained with $\Delta x = 0.05$ and 
 $\Delta t = 0.000125$ for different values of $\cal M$.
}
\begin{center}
\begin{tabular}{| c | c | c | c | c | c | c |}
\hline
\multicolumn{1}{ |c| }{} &
\multicolumn{2}{|c|}{Eqs. (\ref{cgpet3d-1})-(\ref{cgpet3d-3})}&
\multicolumn{2}{ |c| }{TCSM} &\multicolumn{2}{ |c| } {$^{23}$Na}\\
\hline
\multicolumn{1}{ |c| }{${\cal M}$}& 
\multicolumn{1}{ |c| } {$\mu_{+2}$}& 
\multicolumn{1}{ |c| }{$\mu_{-2}$}& 		
\multicolumn{1}{ |c| }{$\mu_{+2}$}& 		
\multicolumn{1}{ |c| }{$\mu_{-2}$}&
\multicolumn{1}{ |c| }{$E$ } & 		
\multicolumn{1}{ |c| }{$E$- TCSM}\\
\hline
0.0&25.3397&25.3397&25.3397&25.3397&15.2216&15.2216 \\ \hline
0.2&25.7025&24.9641&25.7025&24.9641&15.2401&15.2401\\ \hline
0.4&26.0570&24.5702&26.0570&24.5702&15.2956&15.2956\\ \hline
0.6&26.4064&24.1517&26.4064&24.1517&15.3891&15.3891 \\ \hline
0.8&26.7530&23.7013&26.7530&23.7013&15.5216&15.5216 \\ \hline 
1.0&27.0978&23.2103&27.0978&23.2103&15.6949&15.6949 \\ \hline
1.2&27.4410&22.6678&27.4410&22.6678&15.9112&15.9112 \\ \hline
1.4&27.7824&22.0565&27.7824&22.0565&16.1733&16.1733 \\ \hline
1.6&28.1220&21.3449&28.1220&21.3449&16.4854&16.4854 \\ \hline
1.8&28.4597&20.4577&28.4597&20.4577&16.8538&16.8538 \\ \hline
\end{tabular}
\label{table2}
\end{center}
\end{table}

\begin{table}[!h]
\caption{Ground state energies and chemical potential  values of q1D $^{87}$Rb BEC obtained with 
the spin-2 mean-field model and the TCSM. The results have been obtained with $\Delta x = 0.05$
and $\Delta t = 0.000125$ for different values of $\cal M$.}
\begin{center}
\begin{tabular}{| c | c | c | c | c | c |c | c |}
\hline
\multicolumn{1}{ |c| }{}&
\multicolumn{2}{|c|}{ Eqs. (\ref{cgpet3d-1})-(\ref{cgpet3d-3}) }&
\multicolumn{2}{ |c| } {TCSM} &
\multicolumn{2}{ |c| } {$^{87}$Rb}\\
\hline
\multicolumn{1}{ |c| }{${\cal M}$}& 
\multicolumn{1}{ |c| } {$\mu_{+2}$}& 
\multicolumn{1}{ |c| }{$\mu_{-1}$}& 		
\multicolumn{1}{ |c| }{$\mu_{+2}$}& 		
\multicolumn{1}{ |c| }{$\mu_{-1}$}&
\multicolumn{1}{ |c| }{$E$ } & 		
\multicolumn{1}{ |c| }{$E$- TCSM}\\
\hline
0.1&58.0251&57.8798&58.0251&57.8798&34.7691&34.7691 \\ \hline
0.3&58.2121&57.7760&58.2121&57.7760&34.7884&34.7884 \\ \hline
0.5&58.3952&57.6631&58.3952&57.6631&34.8273&34.8273\\ \hline
0.7&58.5766&57.5383&58.5766&57.5383&34.8863&34.8863 \\ \hline
0.9&58.7578&57.3979&58.7578&57.3979&34.9661&34.9661 \\ \hline
1.1&58.9393&57.2376&58.9393&57.2376&35.0680&35.0680 \\ \hline
1.3&59.1209&57.0512&59.1209&57.0512&35.1936&35.1936 \\ \hline
1.5&59.3026&56.8284&59.3026&56.8284&35.3448&35.3448\\ \hline
1.7&59.4841&56.5485&59.4841&56.5485&35.5247&35.5247 \\ \hline
1.9&59.6655&56.1488&59.6655&56.1488&35.7387&35.7387 \\ \hline
\end{tabular}
\label{table3}
\end{center}
\end{table}

\subsubsection*{With SO coupling, $\gamma_x\ne 0$}
\begin{figure}
\centering
\includegraphics[width=1\linewidth]{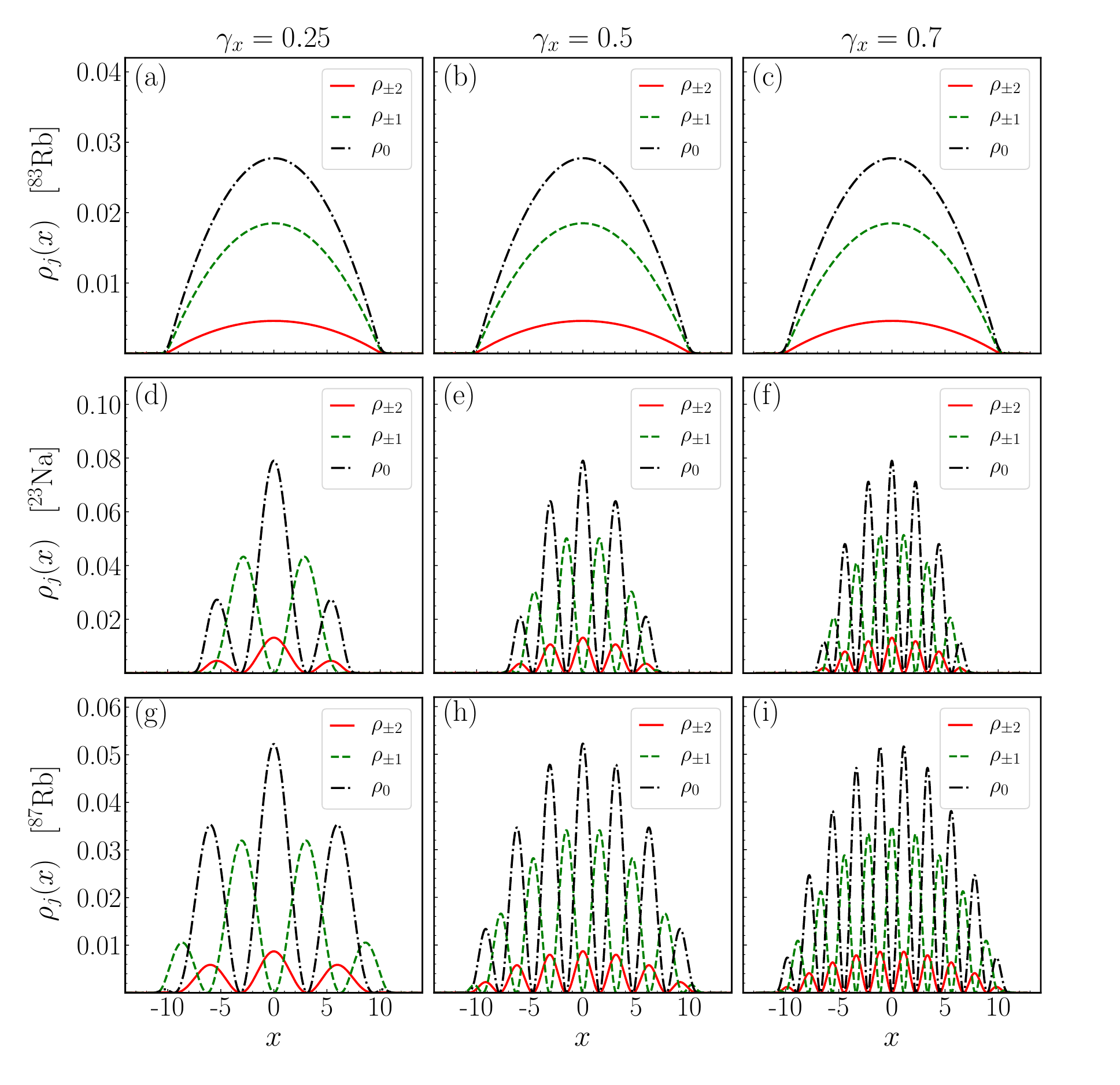}
\caption {(Color online) (a)-(c) Numerically calculated ground state densities
 for an SO-coupled $^{83}$Rb with $\gamma_x = 0.25, 0.5,$ and $0.7$, respectively. 
 The same for $^{23}$Na and $^{87}$Rb are shown in sub-figures (d)-(f) and (g)-(i), 
 respectively.}
\label{fig2}
\end{figure}

In the presence of SO coupling, for $^{83}$Rb, $^{23}$Na and 
$^{87}$Rb, we again consider the same parameters as we have chosen for 
$\gamma_x=0$ in Sec. \ref{q1d}. The values of individual component
chemical potentials and  total energies for $^{83}$Rb, $^{23}$Na, $^{87}$Rb with 
various values of $\gamma_x$ are presented in tables \ref{table5a}, \ref{table5b} and 
\ref{table5c}, respectively. The component densities for three systems with 
 $\gamma_x = 0.25$, 0.5 and 0.7 are illustrated in Fig. \ref{fig2}.
 
\begin{table}
\caption{Ground state energies and chemical potential values of $^{83}$Rb condensates in the 
presence of SO coupling with $\Delta x = 0.05$ , $\Delta t = 0.000125$}. 
\begin{center}
\begin{tabular}{| c | c | c |c | c | c |c |}
\hline
\multicolumn{1}{|c| }{$\gamma_x$}&
\multicolumn{1}{|c|}{$\mu_{2}$}&
\multicolumn{1}{|c| }{$\mu_{1}$}&
\multicolumn{1}{|c| }{$\mu_{0}$}&
\multicolumn{1}{|c| }{$\mu_{-1}$}&
\multicolumn{1}{|c| }{$\mu_{-2}$}&
\multicolumn{1}{|c|}{Energy}\\
\hline
0.1&51.3774&51.3774&51.3774&51.3774&51.3774&30.8296 \\ \hline
0.2&51.3172&51.3172&51.3172&51.3172&51.3172&30.7696 \\ \hline
0.3&51.2169&51.2169&51.2169&51.2169&51.2169&30.6696\\ \hline
0.4&51.0764&51.0764&51.0764&51.0764&51.0764&30.5296\\ \hline
0.5&50.8958&50.8958&50.8958&50.8958&50.8958&30.3496\\ \hline
0.6&50.6751&50.6751&50.6751&50.6751&50.6751&30.1296\\ \hline
0.7&50.4142&50.4142&50.4142&50.4142&50.4142&29.8696 \\ \hline
0.8&50.1131&50.1131&50.1131&50.1131&50.1131&29.5696\\ \hline
0.9&49.7720&49.7720&49.7720&49.7720&49.7720&29.2296\\ \hline
1.0&49.3907&49.3907&49.3907&49.3907&49.3907&28.8496\\ \hline
\end{tabular}
\label{table5a}
\end{center}
\end{table}

\begin{table}
\caption{Ground state energies and chemical potential values of {$^{23}$Na} condensates in the 
presence of SO coupling with $\Delta x = 0.05$ , $\Delta t = 0.000125$}. 
\begin{center}
\begin{tabular}{| c | c | c |c | c | c |c |}
\hline
\multicolumn{1}{|c| }{$\gamma_x$}&
\multicolumn{1}{|c|}{$\mu_{2}$}&
\multicolumn{1}{|c| }{$\mu_{1}$}&
\multicolumn{1}{|c| }{$\mu_{0}$}&
\multicolumn{1}{|c| }{$\mu_{-1}$}&
\multicolumn{1}{|c| }{$\mu_{-2}$}&
\multicolumn{1}{|c|}{Energy}\\
\hline
0.1&25.3192&25.3199&25.3192&25.3199&25.3192&15.2016 \\ \hline
0.2&25.2588&25.2597&25.2588&25.2597&25.2588&15.1416 \\ \hline
0.3&25.1591&25.1593&25.1591&25.1593&25.1591&15.0416 \\ \hline
0.4&25.0186&25.0193&25.0186&25.0193&25.0186&14.9016 \\ \hline
0.5&24.8385&24.8388&24.8385&24.8388&24.8385&14.7216 \\ \hline
0.6&24.6181&24.6185&24.6181&24.6185&24.6181&14.5016 \\ \hline
0.7&24.3576&24.3581&24.3576&24.3581&24.3576&14.2416\\ \hline
0.8&24.0572&24.0575&24.0572&24.0575&24.0572&13.9416\\ \hline
0.9&23.7165&23.7170&23.7165&23.7170&23.7165&13.6016\\ \hline
1.0&23.3360&23.3363&23.3360&23.3363&23.3360&13.2216\\ \hline
\end{tabular}
\label{table5b}
\end{center}
\end{table}

\begin{table}
\caption{Ground state energies and chemical potential values of {$^{87}$Rb} condensates in the 
presence of SO coupling with $\Delta x = 0.05$ , $\Delta t = 0.000125$}. 
\begin{center}
\begin{tabular}{| c | c | c |c | c | c |c |}
\hline
\multicolumn{1}{|c| }{$\gamma_x$}&
\multicolumn{1}{|c|}{$\mu_{2}$}&
\multicolumn{1}{|c| }{$\mu_{1}$}&
\multicolumn{1}{|c| }{$\mu_{0}$}&
\multicolumn{1}{|c| }{$\mu_{-1}$}&
\multicolumn{1}{|c| }{$\mu_{-2}$}&
\multicolumn{1}{|c|}{Energy}\\
\hline
0.1&57.9121&57.9121&57.9120&57.9121&57.9121&34.7484 \\ \hline
0.2&57.8518&57.8516&57.8518&57.8516&57.8518&34.6884 \\ \hline
0.3&57.7513&57.7513&57.7513&57.7513&57.7513&34.5884\\ \hline
0.4&57.6108&57.6108&57.6108&57.6108&57.6108&34.4484\\ \hline
0.5&57.4302&57.4300&57.3402&57.4300&57.4302&34.2684\\ \hline
0.6&57.2092&57.2093&57.2092&57.2093&57.2092&34.0484\\ \hline
0.7&56.9483&56.9482&56.9483&56.9482&56.9483&33.7884\\ \hline
0.8&56.6471&56.6470&56.6471&56.6470&56.6471&33.4884\\ \hline
0.9&56.3058&56.3058&56.3058&56.3058&56.4058&33.1484\\ \hline
1.0&55.9243&55.9243&55.9243&55.9243&55.9243&32.7684\\ \hline
\end{tabular}
\label{table5c}
\end{center}
\end{table}

\subsubsection{Real-time Dynamics}
\begin{figure}
\includegraphics[width=1\linewidth]{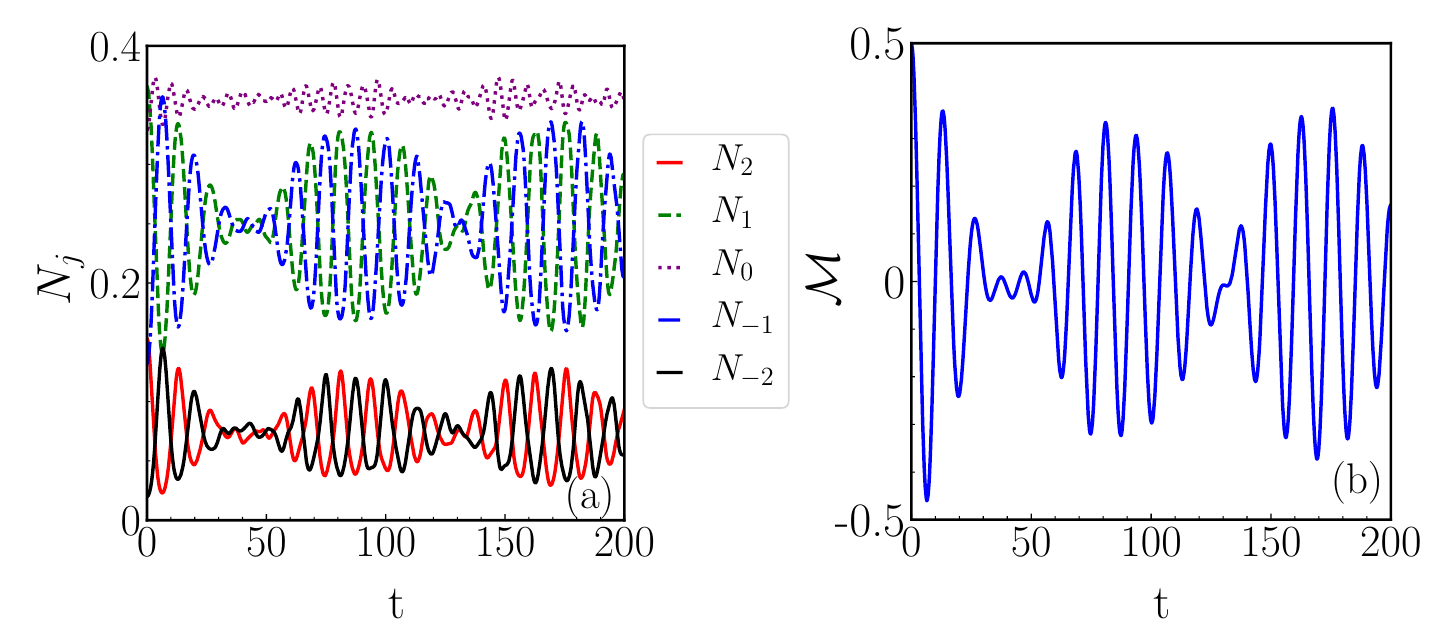}
 \caption {(Color online) (a) Component norms and (b) magnetization as a function of time 
 for $^{83}$Rb with $\tau_0=699.62$, $\tau_1=-1.23$, and $\tau_2=4.90$ with magnetization at $t = 0$
 set to $0.5$.}
\label{fig7}
\end{figure}

We consider q1D $^{83}$Rb BEC with SO coupling strength $\gamma_x = 0.5$ to illustrate
an example of the real-time dynamics which can be investigated with the three programs. We prepare
an initial solution with a fixed magnetization of $0.5$ using imaginary-time propagation. Fixing
the magnetization ensures that the solution obtained with imaginary-time propagation is not the 
ground state; as to obtain the ground state solution, shown in Fig. \ref{fig2}(b), magnetization 
is not fixed in imaginary-time propagation. Hence the solution is expected to show spin-mixing
dynamics when evolved in real time. This is evident from variation of the component norms 
$N_j$'s and magnetization ${\cal M}$ as a function of time $t$ shown in Figs. \ref{fig7}(a) and
\ref{fig7}(b), respectively.

\subsection{Results for q2D spin-2 BECs}
Here also, we consider three cases (a) $^{83}$Rb, (b) $^{23}$Na, and (c) $^{87}$Rb
spin-2 BECs. We consider $10000$ atoms of these systems trapped in q2D trapping potential
with $\omega_x = \omega_y = 2\pi\times 20$ Hz, $\omega_z = 2\pi\times 400$ Hz, and
thus $\alpha_x = \alpha_y = 1$ and $ \alpha_z = 20$. The triplet of dimensionless interaction 
strengths $(\tau_0, \tau_1,\tau_2)$ for these cases are given as
\begin{eqnarray*}
{\rm (a)}\quad (\tau_0, \tau_1,\tau_2) &=& (392.14, -0.67, 2.74),\\
{\rm (b)}\quad (\tau_0, \tau_1, \tau_2) &=& (136.18, 6.76, -7.30),\\
{\rm (c)}\quad (\tau_0,\tau_1, \tau_2) &=& (465.92, 5.55, 0.18).
\end{eqnarray*}
The comparison of ground state energies between full mean-field model and scalar models of spin-2
BEC for all three cases with different values of magnetization is excellent as is reported in
Table \ref{table6}. 

\begin{figure}
 \includegraphics[width=1\linewidth]{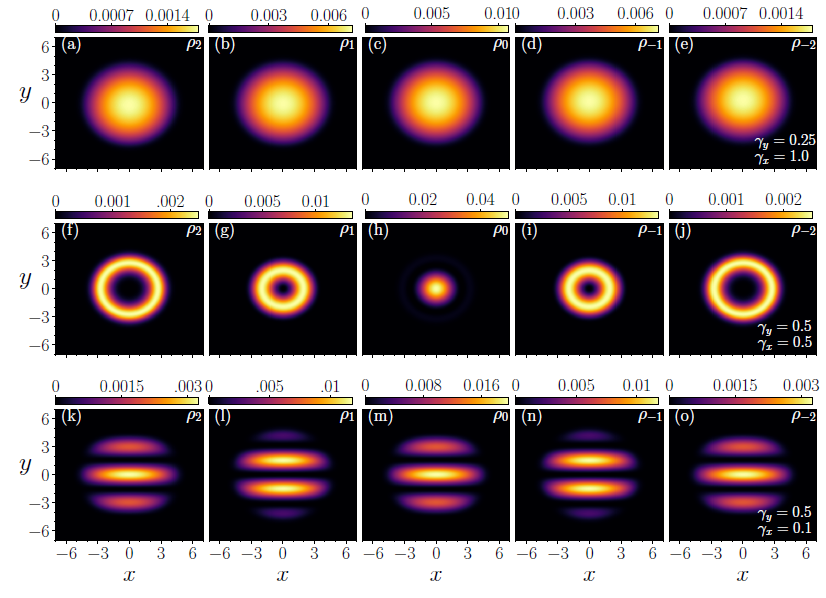}
 \caption {(Color online) (a)-(e) are the ground state component densities for an SO-coupled
 q2D spin-2 BEC with $\tau_0=392.14,\tau_1=-0.69, \tau_2=2.75$, $\gamma_x=1,$ and $\gamma_y=0.25$.
 (f)-(j) and (k)-(o) are the same for the SO-coupled spin-2 BECs with $\tau_0=136.18, \tau_1=6.76, 
 \tau_2=-7.36$, $\gamma_x=0.5, \gamma_y=0.5$ and $\tau_0=465.920, \tau_1=5.55, \tau_2=0.18, \gamma_x=0.1, 
 \gamma_y=0.5$, respectively.}
\label{fig3}
\end{figure}

In the presence of SO coupling, we also illustrate some of the qualitatively distinct numerically
obtained ground state solutions for q2D configurations. We consider interaction parameters' set 
(a) with $\gamma_x=1$ and $\gamma_y=0.25$, set (b) with $\gamma_x=\gamma_y = 0.5$, and set (c)
with $\gamma_x=0.1$ and $\gamma_y=0.5$. The distinct nature of density profiles for the three
cases is evident from Figs. \ref{fig3}(a)-(o). The ground state solution for $^{83}$Rb is a 
plane-wave solution with Gaussian density profiles for all the five components as is shown in Fig 
\ref{fig3}(a)-\ref{fig3}(e). For $^{23}$Na, ground state solution has vortices of winding numbers
$-2,-1,0,+1,$ and $+2$ associated with $m_f= +2,+1,0,-1,$ and $-2$ components, respectively. The
non-zero vorticity associated with components $m_f = \pm1,\pm2$ leads to the zero densities at the
center of these components as is shown in Figs. \ref{fig3}(f)-(j). For $^{87}$Rb, the ground
state has horizontal stripe pattern in the component densities as is shown in Figs. 
\ref{fig3}(k)-(o).

\begin{figure}
\includegraphics[width=1\linewidth]{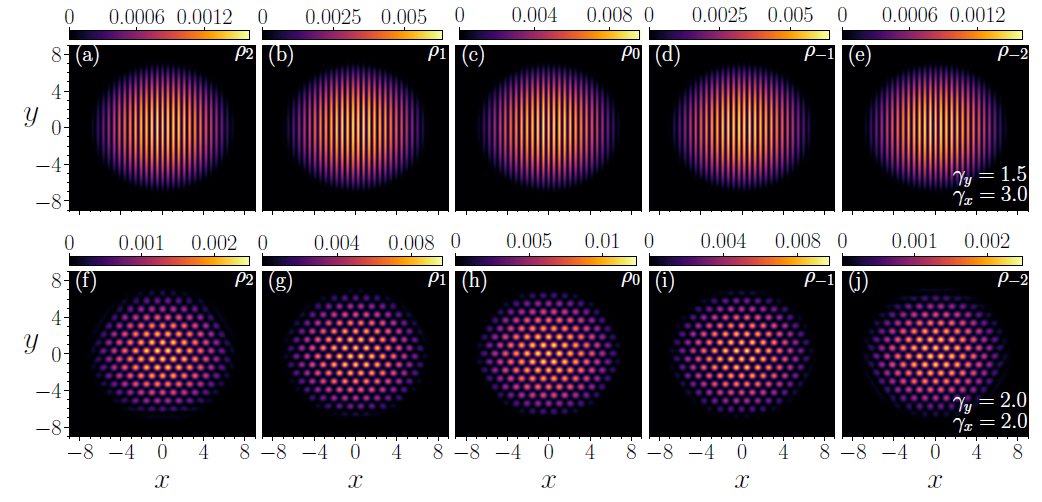}
\caption {(Color online) (a)-(e) are the ground state component densities for an SO-coupled
 q2D spin-2 BEC with $\tau_0=2000, \tau_1=400, \tau_2=-400$, $\gamma_x=3$ and $\gamma_y = 1.5$.
 (f)-(j) are the same for a spin-2 BEC with $\tau_0=2000, \tau_1=40, \tau_2=400$, $\gamma_x=\gamma_y=2$.
 These results are in agreement with Ref. \cite{Xu}}
 \label{fig4}
\end{figure}

We have also confirmed the accuracy of our codes by comparing our results with results reported 
in Ref. \cite{Xu}. As per the parameters considered in Ref. \cite{Xu},
we consider q2D spin-2 BEC firstly with $\tau_0=2000, \tau_1=400$, $\tau_2=-400$, $\gamma_x=3$, $\gamma_y 
=1.5$ and secondly with  $\tau_0=2000, \tau_1=40$, $\tau_2=400$, $\gamma_x = \gamma_y=2$. The ground state
density profiles in these two cases as shown in Figs. \ref{fig4}(a)-(e) and Figs. 
\ref{fig4}(f)-(j), respectively, are in agreement with Ref. \cite{Xu}. For all these
2D results reported in this subsection, we have considered spatial step sizes $\Delta$x =
$\Delta$y = 0.05 and temporal step size $\Delta$t = 0.000125. 

\begin{table}[!h]
\caption{Ground state energies of $^{83}$Rb, $^{23}$Na,
and $^{87}$Rb q2D spin-2 BECs obtained with full mean-field model and
scalar models (SCSM for $^{83}$Rb and TCSM for $^{23}$Na and $^{87}$Rb).
The results have been obtained with $\Delta x = 0.05$ and
$\Delta t = 0.000125$ for different values of $\cal M$.}
\begin{center}
\small\addtolength{\tabcolsep}{-5pt}
\begin{tabular}{| c | c | c | c | c | c | c | c |}
\hline
\multicolumn{1}{ |c| }{}&\multicolumn{2}{|c|}{ $^{83}$Rb}&
\multicolumn{2}{ |c| } {$^{23}$Na} &\multicolumn{2}{ |c| }{$^{87}$Rb }\\
\hline
\multicolumn{1}{ |c| }{${\cal M}$}& 	
\multicolumn{1}{ |c| } {$E$}& 
\multicolumn{1}{ |c| }{$E$ - SCSM} & 		
\multicolumn{1}{ |c| }{$E$}& 		
\multicolumn{1}{ |c| }{$E$- TCSM} &
\multicolumn{1}{ |c| } {$E$}&
\multicolumn{1}{ |c| }{$E$ - TCSM} \\
\hline
0.0&7.5336&7.5336&4.5314&4.5314&8.2229&8.2229\\\hline
0.2&7.5336&7.5336&4.5352&4.5352&8.2247&8.2247\\\hline
0.4&7.5336&7.5336&4.5468&4.5468&8.2300&8.2300\\\hline
0.6&7.5336&7.5336&4.5662&4.5662&8.2388&8.2388\\\hline
0.8&7.5336&7.5336&4.5937&4.5937&8.2512&8.2512\\\hline
1.0&7.5336&7.5336&4.6294&4.6294&8.2673&8.2673\\\hline
1.2&7.5336&7.5336&4.6739&4.6739&8.2872&8.2872\\\hline
1.4&7.5336&7.5336&4.7278&4.7278&8.3112&8.3112\\\hline
1.6&7.5336&7.5336&4.7921&4.7921&8.3396&8.3396\\\hline
1.8&7.5336&7.5336&4.8684&4.8684&8.3729&8.3729\\\hline
\end{tabular}
\label{table6}
\end{center}
\end{table}

\subsection{Results for 3D spin-2 BECs}
\begin{figure}[H]
\begin{center}
\begin{tabular}{lll}
\includegraphics[trim = 0.05cm 0.05cm 0.05cm 0.05cm, clip,width=1.0\linewidth,
clip]{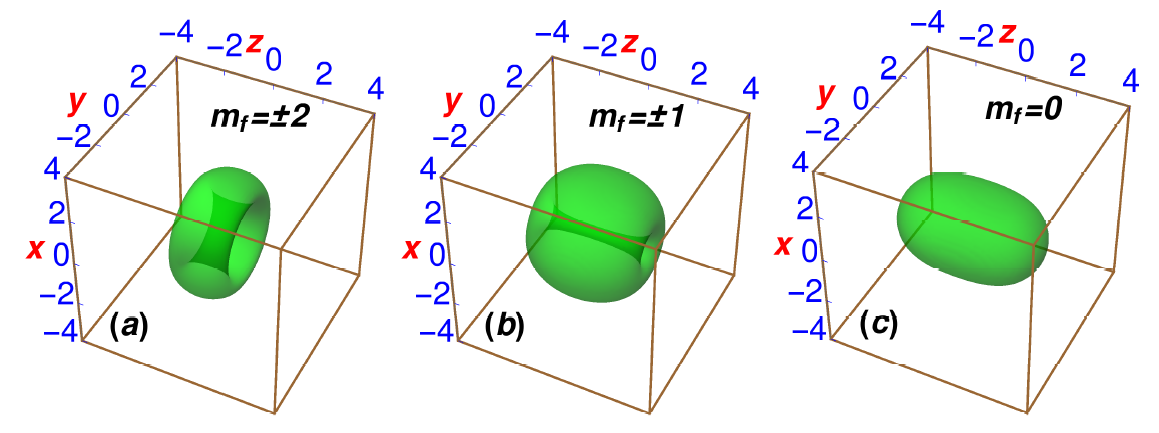}\\
\end{tabular}
\caption{(a)-(c) 3D isosurfaces of
component densities for an SO-coupled $^{23}$Na spin-2 BEC with $\tau_0=76.33, \tau_1=3.79, 
\tau_2=-4.09$ and $\gamma_x=\gamma_y=\gamma_z = 0.5$. The isosurfaces correspond to isodensity value 
of $0.0007$. The solution has been obtained with $\Delta x = 0.1$ and $\Delta t = 0.0005$.}
\label{fig8}
\end{center}
\end{figure}

We consider a $^{23}$Na spin-2 BEC with $\tau_0=76.33, \tau_1=3.79$, $\tau_2=-4.09$, 
$\alpha_x=\alpha_y=\alpha_z =1 $ and spin-orbit coupling strengths 
$\gamma_x=\gamma_y=\gamma_z=0.5$. This set of parameters corresponds to 10000 $^{23}$Na atoms 
trapped in an isotropic trapping potential with $\omega_x=\omega_y=\omega_z = 2\pi\times 20$Hz. 
Here we obtain (-2,-1,0,+1,+2) type of vortex solution and 
corresponding energy is 1.9693. The 3D iso-surfaces corresponding to iso-density value of
of $0.0007$ are shown in Fig. \ref{fig8}.

\section{Summary}
We have provided a set of three OpenMP parallelized FORTRAN 90/95 programs to solve the CGPEs describing an $f=2$ spinor BEC with an anisotropic spin-orbit coupling in q1D, q2D and 3D 
configurations using the time-splitting Fourier spectral method. These codes can be used to simulate
both static and dynamic properties of an SO-coupled spin-2 BEC with a variety of SO couplings
including Rashba, Dresselhaus or a combination of both. We have confirmed the accuracy of the
codes by comparing the the results for ground-state energies, chemical potentials and densities
obtained from the codes with those available in the literature or with the simplified scalar models 
which show an excellent agreement. The test results for OpenMP performance parameters like
speedup and efficiency are very good. With the advent of SO coupling in the spinor BECs, the
present set of codes can be very useful to the researchers working on spin-2 BECs.

\section*{Appendix}
\subsection*{Conservation/Non-conservation of Magnetization}
In the absence of spin-orbit coupling, $\Gamma_{\pm 2}({\bf x},t) =0, \Gamma_{\pm 1}({\bf 
x},t)=0, \Gamma_{0}({\bf x},t)=0$, 

\begin{eqnarray}
\frac{d{\cal M}}{d t} &=& \int \left( 2|\phi_{+2}({\bf x},t)|^2 +|\phi_{+1}({\bf x},t)|^2-|\phi_{-1}({\bf x},t)|^2-2|\phi_{-2}({\bf x},t)|^2\right) d{\bf x},\nonumber\\ 
&=& \int \left( 2\frac{\partial \phi_{+2}}{\partial t}\phi_{+2}^* +  2\phi_{+2}\frac{\partial 
\phi_{+2}^*}{\partial t} + \frac{\partial \phi_{+1}}{\partial t}\phi_{+1}^*+ \phi_{+1}\frac{\partial 
\phi_{+1}^*}{\partial t}- \frac{\partial \phi_{-1}}{\partial t}\phi_{-1}^*\right.\nonumber \\
&& \left.-  \phi_{-1}\frac{\partial \phi_{-1}^*}{\partial t} -2\frac{\partial \phi_{-2}}{\partial 
t}\phi_{-2}^* -  2\phi_{-2}\frac{\partial \phi_{-2}^*}{\partial t}\right) d{\bf x}. \label{magexpr}
\end{eqnarray}
Using Eqs. (\ref{cgpet3d-1})-(\ref{cgpet3d-3}) in Eq. (\ref{magexpr}), we obtain
\begin{eqnarray}
 \frac{d{\cal M}}{d t}&=& -i\tau_1\int \left( F_{-} \phi_{+1} \phi_{+2}^*  -  F_{+} \phi_{+1}^* \phi_{+2} - F_{+} 
 \phi_{-1} \phi_{-2}^* + F_{-} \phi_{-1}^* \phi_{-2} +\sqrt{\frac{3}{2}}F_{-} \phi_{0} 
 \phi_{+1}^*\right.\nonumber\\
 && \left.- \sqrt{\frac{3}{2}}F_{+} \phi_{0}^* \phi_{+1} -\sqrt{\frac{3}{2}}F_{+} \phi_{0}  
 \phi_{-1}^*+\sqrt{\frac{3}{2}}F_{-} \phi_{0}^* \phi_{-1} \right) d{\bf x},\nonumber\\
 &=&-i\tau_1\int \left[ F_{-}\left( \phi_{+1} \phi_{+2}^* +\phi_{-1}^* \phi_{-2} 
 +\sqrt{\frac{3}{2}} \phi_{0}^* \phi_{-1} +\sqrt{\frac{3}{2}}\phi_{0} \phi_{+1}^*  
 \right)\right.\nonumber \\
 && \left.-  F_{+} \left(\phi_{+1}^* \phi_{+2} + \phi_{-1} \phi_{-2}^* +   \sqrt{\frac{3}{2}} \phi_{0}^* \phi_{+1} +\sqrt{\frac{3}{2}} \phi_{0}  \phi_{-1}^*\right)\right] d{\bf x},\nonumber\\
&=& -i\tau_1\int \left( F_{-}\frac{F_{+}}{2} -  F_{+}\frac{F_{-}}{2} \right) d{\bf x}\nonumber \\
 &=&0.
\end{eqnarray}
In the presence of SO coupling, combining Eqs. (\ref{cgpet3d-1})-(\ref{cgpet3d-3}) with Eq. 
(\ref{magexpr}) leads to
\begin{eqnarray}
\frac{d{\cal M}}{d t}  &=& -i \int\left( 2\phi_{+2}^*\Gamma_{+2} + 2 \phi_{+2}\Gamma_{+2}^* 
+\phi_{+1}^*\Gamma_{+1}+\phi_{+1}\Gamma_{+1}^* -2\phi_{-2}^*\Gamma_{-2} + 2 \phi_{-2}\Gamma_{-2}^* 
\right.\nonumber\\
&&\left.+\phi_{-1}^*\Gamma_{-1}+\phi_{-1}\Gamma_{-1}^* \right) d{\bf x}\nonumber\\
 &\ne &0,
\end{eqnarray}
in general. Therefor ${\cal M}$ is conserved in the absence of SO coupling, but not so, in general,
in the presence of SO coupling.
\subsection*{Simultaneous conservation of Norm and Magnetization}
\label{subsection-II-1}
We use imaginary time propagation method, where $t$ is replaced by $-i\tau$ in CGPEs, viz. Eqs. 
(\ref{cgpet3d-1})-(\ref{cgpet3d-3}), to determine the stationary states of the system. Now, as
the imaginary time propagation used to calculate the ground state of the system under the
constraint of fixed norm and magnetization, conserves neither of the two, one needs to 
renormalize the component wavefunctions after each time iteration. This means after each 
imaginary-time step $\delta \tau$, the component wavefunctions are rescaled as
$
\phi_j({\bf x}, \tau + \delta t)  = \sigma_j \phi_j({\bf x}, \tau) \label{renom_sol},
$
where $\sigma_j$'s are renormalization factors.
These renormalization factors $\sigma_j$'s satisfy the following 
relationships among them \cite{gautam2015spontaneous}
\begin{subequations}
\begin{eqnarray}
 \sigma_1 \sigma_{-1}  = \sigma_0^2,\label{norm-1}\\
 \sigma_2 \sigma_{-2}  = \sigma_0^2,\label{norm-2}\\
 \sigma_2 \sigma_{-1}^2  = \sigma_0^3\label{norm-3},
\end{eqnarray}
\end{subequations}
and
\begin{subequations}
\begin{eqnarray}
u^4 N_2+u^3 v N_1+ u^2 v^2 N_0+u v^3 N_{-1}+v^4  N_{-2} = {\cal N},\label{norm-6}\\
2 u^4 N_2+ u^3  v N_1- u v^3  N_{-1}- 2v^4  N_{-2} = {\cal M}\label{norm-7}
\end{eqnarray}
\end{subequations} 
with norm and magnetization, where $u = \sigma_1^2$ and 
$v = \sigma_0^2$ and $N_j = \int |\phi_j({\bf x},\tau)|^2 d{\bf x}$ are 
the component norms at (imaginary) time $\tau$.
In the present work, we solve Eqs. (\ref{norm-6})-(\ref{norm-7}) using
Newton-Raphson method after each iteration in imaginary time. The $\sigma_1$
and $\sigma_0$ so obtained can be substituted back in 
Eqs. (\ref{norm-1})-(\ref{norm-3}) to determine the remaining renormalization
factors $\sigma$'s. The simultaneous fixing of norm and magnetization is only 
implemented in the absence of SO coupling.


\begin{thebibliography}{1}
\bibitem{stamper1998optical}
 D.M. Stamper-Kurn, M.R. Andrews, A.P. Chikkatur, S. Inouye, H.-J. 
 Miesner, J. Stenger, and W. Ketterle,
 Phys. Rev. Lett. 80 (1998) 2027.
\bibitem{spin2-phases-ciobanu}
 C.V. Ciobanu, S.-K. Yip, and T.-L. Ho
 Phys. Rev. A 61 (2000) 033607.
\bibitem{spin2-phases-ueda}
 M. Ueda and M. Koashi,
 Phys. Rev. A 65 (2002) 063602.
\bibitem{chang2004observation}
 M.-S. Chang, C.D. Hamley, M.D. Barrett, J.A. Sauer, K.M. Fortier, 
 W. Zhang, L. You, and M.S. Chapman,
 Phys. Rev. Lett. 92 (2004) 140403 
\bibitem{schmaljohann2004dynamics}
 H. Schmaljohann, M.Erhard, J.  Kronj{\"a}ger, M. Kottke, S. Van Staa,
 L. Cacciapuoti, JJ. Arlt, K. Bongs and K. Sengstock,
 Phys. Rev. Lett. 92 (2004) 040402.
\bibitem{kuwamoto2004magnetic}	
 T.Kuwamoto,  K. Araki and T.Hirano,
 Phys. Rev. A. 69 (2004) 063604.
\bibitem{widera2005coherent}
 A. Widera, F. Gerbier, S. Fölling, T. Gericke,
 O. Mandel, and I. Bloch, 
 Phys. Rev. Lett. 95 (2005) 190405.
\bibitem{Kawaguchi-phases}  
 Y. Kawaguchi and M. Ueda
 Phys. Rev. A 84, (2011) 053616.
\bibitem{widera2006precision}
 A. Widera, F. Gerbier, S. Fölling, T. Gericke, O.  Mandel and I.  Bloch,
 New Journal of Physics 8 (2006) 152.
\bibitem{lin2011spin}
 Y.-J. Lin, K. Jim{\'e}nez-Garc{\'\i}a, and I.B. Spielman,
 Nature 471 (2011) 83;
 D.L. Campbell, R.M. Price, A. Putra, A. Vald\'es-Curiel, 
 D. Trypogeorgos, and I.B. Spielman, 
 Nature Communications 7 (2016) 10897;
 X. Luo, L. Wu, J. Chen, Q. Guan, K. Gao, Zhi-Fang Xu, L. You, and R. Wang,
 Scientific Reports 6 (2016) 18983. 
\bibitem{Xu}
 Z.F. Xu, R. L$\ddot{\rm u}$, and L. You
 Phys. Rev. A 83 (2011) 053602.
\bibitem{Kawakami}
 T. Kawakami, T. Mizushima, and K. Machida
 Phys. Rev. A 84 (2011) 011607(R);
 Z. F. Xu, Y. Kawaguchi, L. You, and M. Ueda,
 Phys. Rev. A 86 (2012) 033628.
\bibitem{soc-proposals}
 B. M. Anderson,I. B. Spielman, and G. Juzeli\={u}nas,
 Phys. Rev. Lett. 111 (2013) 125301;
 Z.-F. Xu, L. You, and M. Ueda,
 Phys. Rev. Rev. A 87 (2013) 063634.
\bibitem{kawaguchi2012spinor}
 Y. Kawaguchi, M.  Ueda,
 Physics Reports 520 (2012) 253-381.
\bibitem{spin1}
 H. Wang,
 Int. J. Comp. Math. 84 (2007) 925;
 W. Bao and F.Y. Lim.,
 SIAM Journal on Scientific Computing 30 (2008) 1925;
 W. Bao, I.-L. Chern, and Y. Zhang,
 J. Comp. Phys. 253 (2013) 189.
\bibitem{spin2-wang11}
 H. Wang, J. Comp. Phys. 230 (2011) 6165.
\bibitem{spin2-wang14}
 H. Wang, J. Comp. Phys. 274 (2014) 473.
\bibitem{spin-1/2-soc}
 H. Wang and Z. Xu,
Comp. Phys. Comm. 185 (2014) 2803.
\bibitem{spin1-soc}
 P. Kaur, A. Roy, and  S. Gautam,
 Comp. Phys. Comm. 259 (2021) 107671.
\bibitem{bychkov1984oscillatory}
 Y.A. Bychkov and E.I. Rashba,
 J. Phys. C: Solid state physics 17 (1984) 6039.	
\bibitem{gautam2015spontaneous}
 S. Gautam and S. K. Adhikari, 
 Phys. Rev. A 91 (2015) 013624.	
\bibitem {gautam2015analytic}
 S. Gautam and S. K. Adhikari,
 Phys. Rev. A 92 (2015) 023616.	
\bibitem{Adhikari}
 L.E. Young-S., P. Muruganandam, S.K. Adhikari, 
 V. Loncar, D. Vudragovic, A. Balaz
 Comput. Phys. Commun. 220 (2017) 503;
 V. Loncar, L.E. Young-S., S. Skrbic, 
 P. Muruganandam, S.K. Adhikari, Antun Balaz
 Comput. Phys. Commun. 209 (2016) 190;
 L.E. Young-S., D. Vudragovic, P. Muruganandam, 
 S.K. Adhikari, A. Balaz
 Comput. Phys. Commun. 204 (2016) 209;
 B. Sataric, V. Slavnic, A. Belic,
 A. Balaz, P. Muruganandam, S.K. Adhikari
 Comput. Phys. Commun. 200 (2016) 411;
 V. Loncar, A. Balaz, A. Bogojevic,
 S. Skrbic, P. Muruganandam, S.K. Adhikari
 Comput. Phys. Commun. 200 (2016) 406;
 D. Vudragovic, I. Vidanovic, A. Balaz,
 P. Muruganandam, S.K. Adhikari
 Comput. Phys. Commun. 183 (2012) 2021;
 X. Antoine and R. Duboscq
 Comput. Phys. Commun. 185 (2014) 2969;
 X. Antoine and R. Duboscq
 Comput. Phys. Commun. 193 (2015) 95;
 {\v Z}. Marojevi{\'c} and E. G{\" o}kl{\" u} and 
 Claus L{\"a}mmerzahl Comput. Phys. Commun. 202 (2016) 216
\bibitem{chang2005gauss}
 S.-M. Chang, W.-W. Lin, and S.-F. Shieh,
 J. Comp. Phys. 202 (2005) 367;
 W. Bao and J. Shen,
 SIAM Journal on Scientific Computing 26 (2005) 2010.
\bibitem{ravisankar2020spin}
 R. Ravisankar, D. Vudragovi\'c, P. Muruganandam, A. Bala\v{z}, S.K.Adhikari,
 Comp. Phys. Comm. 259 (2021) 107657.
\bibitem{fftw}
 http://www.fftw.org/
\end{thebibliography}
\end{document}